\documentclass[letterpaper, twocolumn, pra]{revtex4-2}
\usepackage{graphicx, amsmath, amssymb, bm, titlesec, mathtools, braket, natbib, dutchcal, xcolor, upgreek, soul}

\begin{document}
	
	\title{Frequency-bin interferometry for reconstructing electric fields with low intensity}
	
	\author{Abhinandan Bhattacharjee$^{\dagger,*}$, Laura Serino$^{\dagger}$, Patrick Folge$^{\dagger}$, Benjamin Brecht, and Christine Silberhorn}
	\affiliation{Integrated Quantum Optics Group, Institute for Photonic Quantum Systems (PhoQS), Paderborn University, Warburger Straße 100, 33098 Paderborn, Germany}
	\affiliation{$^{\dagger}$These authors contributed equally to this work.\\
		$^{*}$abhib@mail.uni-paderborn.de}
	
\begin{abstract}
	Ultrafast single-photon pulses with tailored time-frequency properties are highly attractive for quantum information science, offering high-dimensional encoding and compatibility with integrated optics platforms. However, accurate characterization of such pulses, including spectral coherence, remains challenging because current methods require substantial experimental resources and complex reconstruction algorithms. Here, we introduce frequency-bin interferometry for reconstructing electric fields with low intensity (FIREFLY), a technique that directly provides spectral amplitude, phase, and coherence profiles of single-photon pulses without requiring intensive reconstruction algorithms. Our approach measures the two-point spectral correlation function of the pulse by interfering its different frequency components using a quantum pulse gate (QPG) driven by a reference pump pulse. We demonstrate its compatibility with quantum light by characterizing partially coherent pulses generated by a type-0 parametric down-conversion process. We also overcome this requirement of a known pump pulse by introducing spectral shear into our interferometric scheme using a multi-output QPG (mQPG). This enables simultaneous characterization of a single-photon-level input pulse alongside an unknown pump pulse. Notably, our method achieves theory-experiment similarity above 95\% across all retrieved profiles, which demonstrates the reliability of this scheme for quantum information applications based on time-frequency encodings. 
\end{abstract} 

\maketitle
\section{Introduction}

The rapid growth in photonic quantum technology \cite{o2009photonic, flamini2018photonic, pelucchi2022potential} has highlighted the importance of single-photon ultrafast pulses \cite{brechtprx2015, ansari2018optica, karpinski2021control}. The {time-frequency} (TF) properties, or mode structure,  of such pulses serve as a powerful resource for quantum information applications, including quantum computing \cite{humphreys2013linear,menicucci2008prl}, quantum communication \cite{bouchard2022prx,bouchard2021achieving}, and quantum metrology \cite{mukamel2020JOPB, donohue2018prl, giovannetti2011advances, barbieri2022prxquantum}. The high-dimensionality, compatibility with integrated optics platforms, and resilience over long-distance propagation further enhance the suitability of TF mode structure for scalable quantum technologies. {Harnessing} these advantages requires precise and complete characterization of the TF properties, namely spectral amplitude, phase, and coherence information. 
Moreover, any quantum information applications demand complete information on single-photon states, hence the knowledge of spectral coherence (or purity) is particularly crucial. {However}, characterizing spectral coherence and phase profiles poses challenges for single-photon pulses.

Conventional pulse characterization techniques such as frequency-resolved optical gating (FROG) \cite{Trebino:93,delong1994josab}, spectral phase interferometry for direct electric-field reconstruction (SPIDER) \cite{walmsley1996josab,iaconis1998optlett,londero2003jmo}, and several of their extensions \cite{fittinghoff1996optlett, gallmann2001optlett, kacprowicz2008complete, reid2000optlett, stibenz2005optlett, pasquazi2011sub, bourassin2015partially} encounter difficulties with low-light level pulses because the nonlinear processes involved in these techniques demand bright input pulses. Alternative methods, such as electro-optic shearing interferometry (EOSI) \cite{davis2018pra, davis2018prl, golestani2022prl}, two-photon spectral interferometry (TPSI) \cite{Thiel:20, lipka2021prl, thekkadath2022prl}, homodyne tomography (HT) \cite{qin2015complete}, Hong-Ou-Mandel interferometry (HOMI) \cite{wasilewski2007prl}, and chronocyclic $Q-$function tomography \cite{bhattacharjee2024pulse} have been used for single-photon pulse characterization. TPSI, HT, and HOMI can characterize pulses with arbitrary spectral coherence, {but} they require a spatially mode-matched and spectrally known reference pulse and coincidence detection. On the other hand, EOSI and $Q-$function tomography have {only} been demonstrated for perfectly coherent pulses. All {of} these techniques rely on computationally intensive reconstruction algorithms, {and often} struggle to detect complex spectral features such as phase jumps. Moreover, their implementations {require} resources such as modulators for temporal phase manipulation, interferometric stability, long measurement time, and spectrally resolved coincidence measurements. 
\begin{figure*}[t!]
	\centering
	\includegraphics[width=0.6\linewidth]{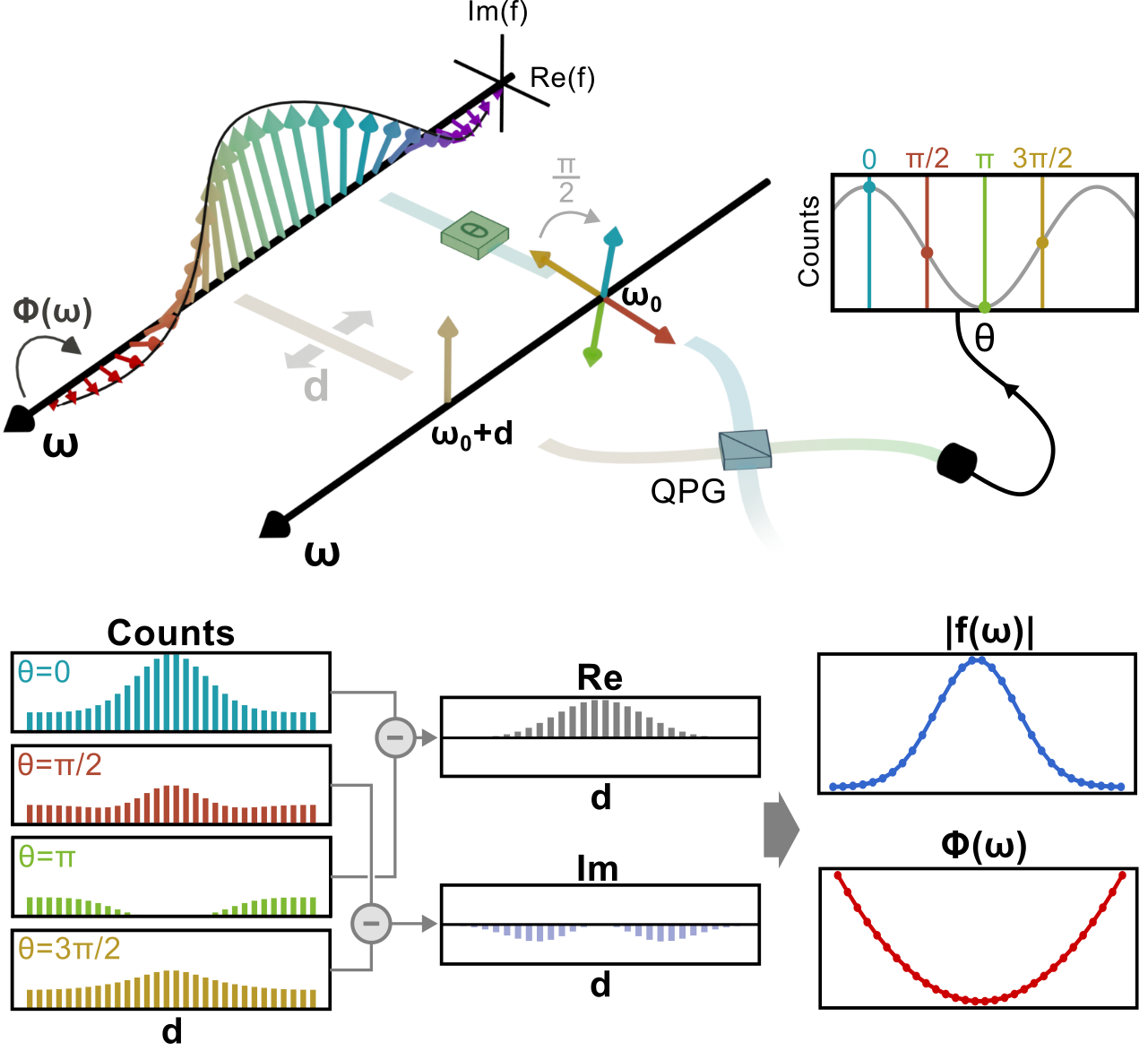}
	\caption{Conceptual illustration of our {approach} for characterizing single-photon pulses by measuring their two-point spectral correlation (TPSC) function. In the illustration, the orientation of the vector represents the spectral phase, while its length corresponds to the spectral amplitude at a specific frequency. {Our approach} interferes different spectral components of the pulse under test using and measure the interfering count distribution as a function of their separation at different relative phase. Using these count distributions, we directly obtain the real and imaginary parts of the TPSC function, which provide the spectral amplitude and phase profiles. 
	}\label{fig1}
\end{figure*} 

In this article, we present frequency-bin interferometry for reconstructing electric fields with low intensity (FIREFLY), a direct, reference-free, and quantum-compatible pulse characterization technique adapted from optical coherence theory \cite{mandel1995optical}. FIREFLY measures the two-point spectral correlation (TPSC) function of single-photon-level pulses and directly yields the spectral amplitude, phase, and coherence profiles without employing a complex reconstruction algorithm. Similar approaches have been widely used for characterizing mode structure and coherence information in the spatial domain \cite{kulkarni2017single, iaconis1996direct, rezvani2017wide, bhattacharjee2018apl}. Our scheme {harnesses} a quantum pulse gate (QPG) \cite{Ansaripra2017, allgaier2017QST, ansari2018prl, gil2021optical, ansari2020optexp}, an integrated dispersion engineered sum-frequency conversion process, to interfere different spectral components of the input pulse and extract the TPSC function by measuring the output counts. First, we demonstrate high-quality characterization of spectrally perfectly coherent pulses from a mode-locked laser attenuated to the single-photon-level. Subsequently, we use FIREFLY to characterize spectrally partially coherent quantum pulses generated by a spectrally multi-mode type-0 parametric down-conversion (PDC) process, highlighting its compatibility with quantum light.  

In the first implementation, we assume access to a characterized pump pulse, which serves as a reference pulse. We then eliminate this requirement by extending FIREFLY to a multi-output QPG (mQPG) \cite{serino2023prx} with multiple sum-frequency conversion output channels. 
Each output channel provides access to shifted relative phase profiles between input and pump, which we combine to reconstruct both pulses, demonstrating for the first time   
the simultaneous characterization of a single-photon level input pulse alongside an unknown bright pump pulse. We further showcase the practicality of this scheme by characterizing both pulses at the low light level without requiring a reference.

\section{Theory}
\subsection{Concept}
{We consider a single-photon ultrafast pulsed field that is single-mode in both polarization and spatial degrees of freedom, with arbitrary spectral coherence. The complete TF properties of such a state are described by the two-point spectral correlation (TPSC) function $W(\omega_1,\omega_2)$, which quantifies the correlation between the complex spectral amplitudes at frequencies $\omega_1$ and $\omega_2$, and is defined as}
\begin{equation}
	W(\omega_1,\omega_2) = \langle f^*(\omega_1)f(\omega_2)\rangle_e,\label{W}
\end{equation}
where $f(\omega)$ is the complex spectral amplitude of the electric field at frequency $\omega$, and $\langle\rangle_e$ represents the ensemble average. {The TPSC function serves as the spectral analog of a density matrix.} 

For spectrally perfectly coherent pulses, $W(\omega_1,\omega_2)= f^*(\omega_1)f(\omega_2)$, {and} $f(\omega)=|f(\omega)|e^{i\phi(\omega)}$ completely characterizes the TF mode structure. Here, $|f(\omega)|$ and $\phi(\omega)$ represent the spectral amplitude and phase profiles, respectively. For partially coherent pulses, $W(\omega_1,\omega_2)$ yields the spectral coherence profile as a function of the frequency separation $\Delta\omega\equiv\omega_1-\omega_2$. This approach enables the complete characterization of single-photon pulses, regardless of their spectral coherence property.%
%
\begin{figure*}[t!]
	\includegraphics[scale=1.0]{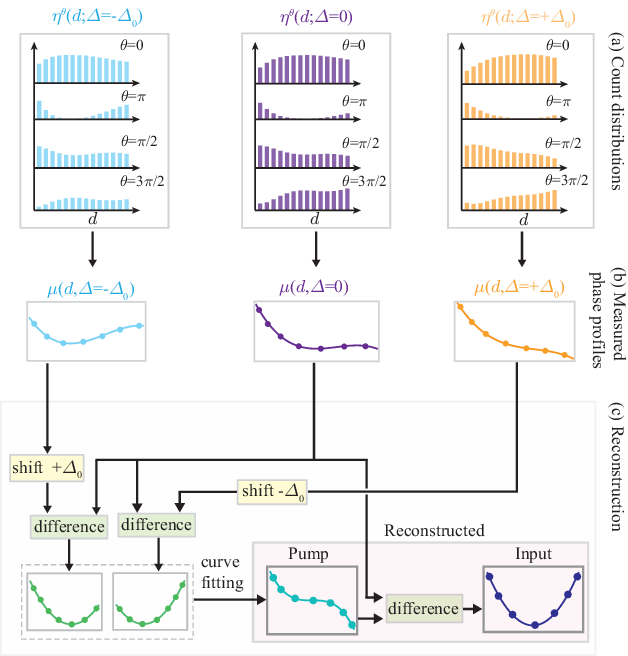}
	\caption{Steps for characterizing the spectral phase profiles of both input and pump pulses using an mQPG. (a) Count distributions for different shearing $\Delta$ (corresponding to different mQPG output channels). (b) Phase profiles $\mu(d, \Delta)$ corresponding to different shearing $\Delta$ obtained from the above count distributions. (c) Reconstruction steps for extracting the input and pump phase profiles from $\mu(d,\Delta)$ profiles.
	}\label{fig2}
\end{figure*} 
Characterizing both spectral phase and coherence typically requires interfering different spectral components of a pulse, a challenging task for single-photon states. We achieve this frequency interference using a QPG, a dispersion-engineered sum-frequency conversion process realized in an integrated waveguide \cite{Ansaripra2017, allgaier2017QST, ansari2018prl, gil2021optical, ansari2020optexp}. Traditionally, this device has been used as a beam splitter for TF modes, selected by spectral shaping of the pump pulse. In this approach, we use the QPG to enable interference between input frequencies by shaping the pump as a superposition of spectral bins. By measuring the up-converted photon counts at the QPG output, we extract the TPSC function, which directly yields the spectral phase and coherence information without requiring reconstruction algorithms. 
	
This first scheme can directly characterize a single-photon input pulse assuming prior knowledge of the pump pulse. In the next step, we eliminate this requirement, and we characterize both input and pump pulses by implementing the same scheme on an mQPG \cite{serino2023prx}. Each output channel of the mQPG applies the equivalent of a well-defined spectral shearing between pump and input frequencies. By measuring photon counts at all the output channels, we reconstruct the spectral phase profile of the pump, and we combine it with the output count data to characterize the input pulse. Our approach harnesses the QPG and mQPG as powerful tools for direct and reference-free characterization of ultrafast single-photon pulses.

\subsection{Measurement scheme: general idea and implementation}\label{TPSC function measurement}
The proposed scheme for measuring the TPSC function is illustrated in Fig.~\ref{fig1}. In this scheme, we use a QPG to interfere complex spectral amplitudes $f(\omega_0)$ and $f(\omega_0+d)$ at frequencies $\omega_0$ and $\omega_0+d$, respectively, with a controlled additional relative phase $\theta$. The resulting interference count distribution $\eta^{\theta}(d)$ at the output of the QPG as a function of $d$ and $\theta$ is given by   
\begin{equation}
	\eta^{\theta}(d) = \lvert{f(\omega_0)+f(\omega_0+d)e^{i\theta}}\rvert^2.\label{I-qpg_1}
\end{equation}
This can be rewritten in terms of the TPSC function $W(\omega_0,\omega_0+d)$ as
\begin{multline}
	\eta^{\theta}(d) = |f(\omega_0)|^2 + |f(\omega_0+d)|^2 +2\mathrm{Re}[W(\omega_0,\omega_0+d)]\cos\theta \\ 
	+2\mathrm{Im}[W(\omega_0,\omega_0+d)]\sin\theta,\label{I-qpg_2}
\end{multline}
where $\mathrm{Re}[W(\omega_0,\omega_0+d)]$ and $\mathrm{Im}[W(\omega_0,\omega_0+d)]$ are the real and imaginary parts of $W(\omega_0,\omega_0+d)$ respectively. Therefore, for each separation $d$, we measure $\eta^{\theta}(d)$ at $\theta\in\{0,\frac{\pi}{2},\pi,\frac{3\pi}{2}\}$ and then extract $W(\omega_0,\omega_0+d)$ using
\begin{multline}
	W(\omega_0,\omega_0+d) \propto \left[\eta^{\theta=0}(d) - \eta^{\theta=\pi}(d)\right] \\ 
	+i \left[\eta^{\theta=\frac{\pi}{2}}(d) - \eta^{\theta=\frac{3\pi}{2}}(d)\right].\label{I-qpg_3}
\end{multline}
By varying $d$, this method enables direct measurement of the TPSC function $W(\omega_0,\omega_0+d)$, as shown in Fig.~\ref{fig1}. The meaning of this measured TPSC function $W(\omega_0,\omega_0+d)$ relies on the spectral coherence property of the pulse. For spectrally perfectly coherent pulses, $|W(\omega_0,\omega_0+d)|$ and Arg$[W(\omega_0,\omega_0+d)]$ yield the spectral amplitude and phase profiles, respectively. On the other hand, for spectrally partially coherent pulses, $W(\omega_0,\omega_0+d)$ represents the spectral coherence profile. 
 
In this first scheme, prior knowledge of the spectral phase profile of the pump pulse is still necessary, as it directly transfers to the relative phase between the two interfering spectral bins. In Eq.~(\ref{I-qpg_1}), we assume that the pump phase is uniform. However, in realistic experimental conditions, the pump pulse often has a spectral phase profile, which modifies the count distribution $\eta^{\theta}(d)$ as
\begin{equation}
	\eta^{\theta}(d) = \lvert{f(\omega_0)+f(\omega_0+d)e^{i\theta}e^{i\alpha(\omega_p^{(0)}-d)}}\rvert^2,\label{I-qpg_4}
\end{equation}
where $\omega_p^{(0)}$ is the pump frequency corresponding to the input frequency $\omega_0$, and $\alpha(\omega_p^{(0)}-d)$ is the unknown spectral phase introduced by the QPG pump. For a known $\alpha(\omega_p^{(0)}-d)$ profile, we can directly subtract it from the retrieved phase profile to extract the input spectral phase. In the more general case, where $\alpha(\omega_p^{(0)}-d)$ is unknown, we would need to characterize it along with the input phase profile. This could be achieved by applying a relative spectral shear $\Delta$ between pump and input frequencies and comparing count distributions for different shearing values; however, shearing is typically quite complex to implement experimentally.

Instead, we extend our scheme to simultaneously reconstruct both input and pump phase profiles by harnessing an mQPG. An output channel with a spectral shift of $\Delta$ relative to the central one interferes frequencies $\omega_0+\Delta$ and $\omega_0+\Delta+d$ in correspondence of the same pump frequencies in Eq.~(\ref{I-qpg_4}), effectively applying the equivalent of shearing between the two pulses. The resulting interference count distribution at the mQPG output is 

\begin{equation}
	\eta^{\theta}(d;\Delta) = \lvert{f(\omega_0+\Delta)+f(\omega_0+d+\Delta)e^{i\theta}e^{i\alpha(\omega_p^{(0)}-d)}}\rvert^2.\label{I-qpg_5}
\end{equation}
In the experiment, we use a three output mQPG with $\Delta\in\{-\Delta_0,0,+\Delta_0\}$. By recording $\eta^{\theta}(d;\Delta)$ at $\theta\in\{0,\frac{\pi}{2},\pi,\frac{3\pi}{2}\}$ (Fig.~\ref{fig2}(a)), we reconstruct the phase profile $\mu(d;\Delta)$
\begin{equation}
	\begin{aligned}
		\mu(d;\Delta) &= \mathrm{Arg}\left[\frac{\eta^{\theta=0}(d;\Delta)-\eta^{\theta=\pi}(d;\Delta)}{\eta^{\theta=\frac{\pi}{2}}(d;\Delta)-\eta^{\theta=\frac{3\pi}{2}}(d;\Delta)}\right]  \\
		&= \phi(\omega_0+d+\Delta)-\alpha(\omega^{(p)}_0-d)-\phi(\omega_0+\Delta).\label{mqpg1-ph}
	\end{aligned}
\end{equation}
This expression relates the measured phase profile $\mu(d;\Delta)$ to both input and pump spectral phase profiles. If the pump spectral phase profile $\alpha(\omega^{(p)}_0-d)$ is known, $\mu(d;\Delta)$ directly provides the input phase profile $\phi(\omega_0+d+\Delta)$. By analyzing $\mu(d;\Delta)$ profiles (Fig.~\ref{fig2}(b)), we use a straightforward algorithm (Fig.~\ref{fig2}(c)) to extract the pump spectral phase with the only assumption that it follows a polynomial function, which is commonly used in pulse characterization experiments. Once the pump phase is determined, we subtract it from the measured $\mu(d;\Delta)$ to obtain the input phase profile.
\begin{figure*}[p]
	\centering
	\includegraphics[scale=1]{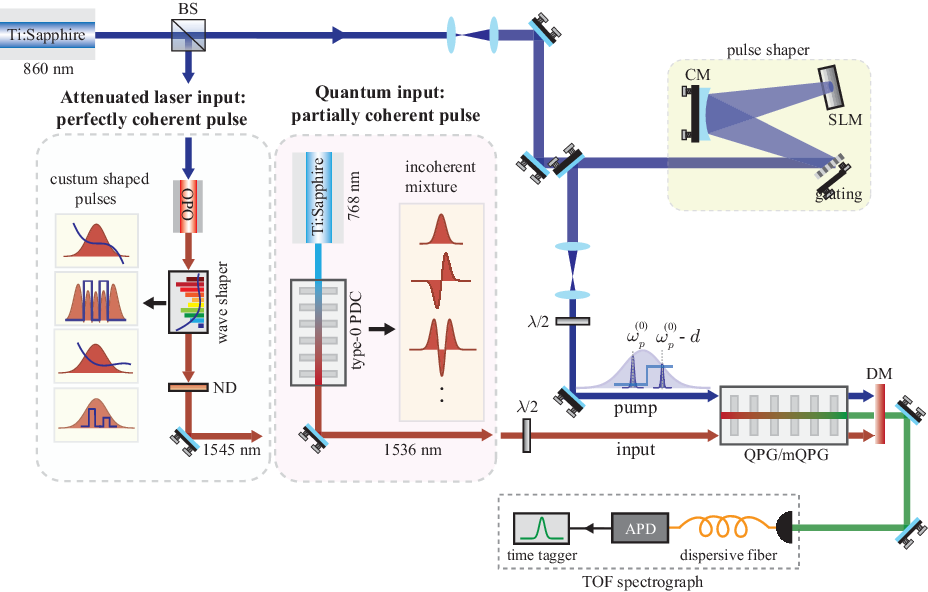}
	\caption{Schematic of the experimental setup. A Ti:Sapphire laser (860 nm) pumps an OPO process to generate spectrally perfectly coherent telecom input pulses (1545 nm). A commercial waveshaper applies custom-designed spectral phase and amplitude profiles to the telecom, which we want to characterize. An ND filter attenuates the input to single-photon-level. To apply this method on a quantum light source, we replace the attenuated OPO pulses by an integrated type-0 PDC. The remaining 860 nm pulse is used as a pump, shaped by an in-house-built pulse shaper. Both input and pump pulses are sent to either the QPG or mQPG waveguide to characterize only the input or both pulses. A time-of-flight spectrograph detects the output pulses. BS: beam splitter, SLM: spatial light modulator, DM: dichroic mirror, CM: cylindrical mirror, OPO: optical parametric oscillator, PDC: parametric down-conversion.}\label{fig3}
\end{figure*}

This approach eliminates the need for an iterative algorithm by replacing it with simple curve fitting of the pump profile, allowing for direct and straightforward reconstruction of the spectral phase profiles of both pulses. Moreover, this method also enables the retrieval of the spectral amplitude profiles (see the supplementary material). 
\section{Experiment and Results}
\subsection{Characterization of spectrally perfectly and partially coherent pulses}
Figure~\ref{fig3} shows the schematic of the experimental setup. A Ti:Sapphire laser, centered at 860 nm, drives an optical parametric oscillator (OPO) to generate spectrally coherent input pulses at 1545 nm. These pulses are attenuated to a mean photon number of 0.1 per pulse using a neutral density (ND) filter, which allows them to be approximated as single-photon pulses. The spectral amplitude and phase profiles are customized using a waveshaper, while the same Ti:Sapphire laser also serves as the pump for the QPG. Additional details are given in the Appendix Sec.~\ref{exp}.

\begin{figure*}[p]
	\centering
	\includegraphics[scale=1.0]{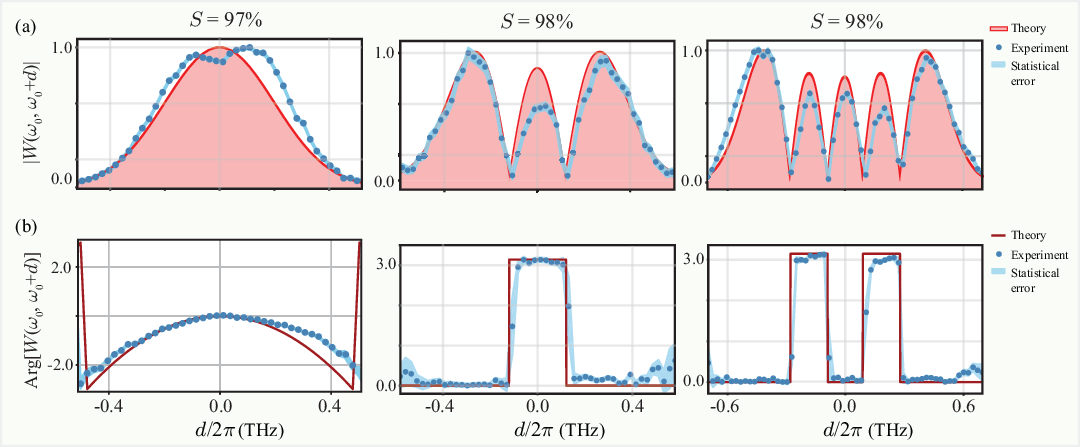}
	\caption{Characterization of single-photon level spectrally perfectly coherent pulses. (a) Measured and theoretical $|W(\omega_0, \omega_0+d)|$ profiles, which represent the functional form of spectral amplitude profile. The maximum of $|W(\omega_0, \omega_0+d)|$ is normalized to 1. (b) Measured and theoretical Arg$[W(\omega_0,\omega_0+d)]$ profiles, which represent the spectral phase profile.}\label{fig4}
\end{figure*}

We first characterize these single-photon perfectly coherent pulses. Figures~\ref{fig4}(a) and ~\ref{fig4}(b) show the measured $|W(\omega_0,\omega_0+d)|$ and Arg$\left[W(\omega_0, \omega_0+d)\right]$, respectively, along with their theoretical predictions for different input pulses. In the supplementary material, we present the step-by-step reconstruction of $|W(\omega_0,\omega_0+d)|$ and Arg$\left[W(\omega_0, \omega_0+d)\right]$ from the measured count distributions. 

For a quantitative comparison, we evaluate the similarity $S$ between theoretical and experimental profiles. We find that $S$ exceeds 97\% for each experimental profile, indicating an excellent agreement with the theory. The high signal to noise ratio of the recorded count distribution ensures almost negligible statistical errors in the reconstructed spectral amplitude and phase profiles. The small deviations in the reconstructed profiles from the expected profiles can be attributed to the experimental imperfections, such as the imperfections in shaping and finite width of pump frequency bins, which effectively lowers the measurement resolution. {Each single-photon photon pulse characterization takes a measurement time of 25 minutes,} though this can be reduced to just a few seconds for bright input pulses. This demonstration highlights the ability of our scheme to characterize spectrally coherent pulses with high precision, including challenging features like phase jumps, a challenging task for existing schemes. 

\begin{figure}[t!]
	\includegraphics[scale=1]{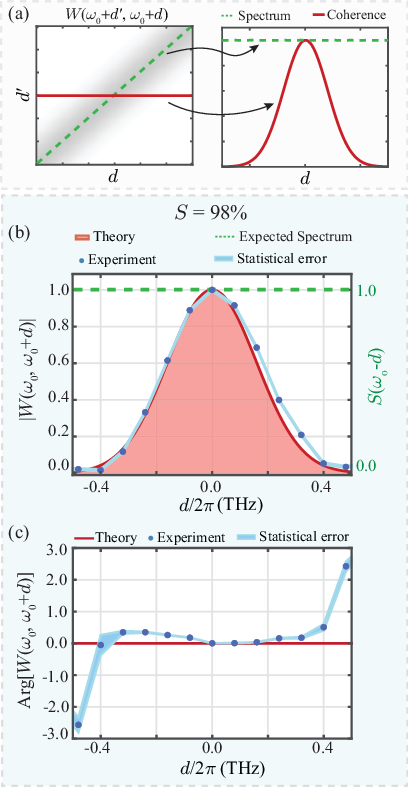}
	\caption{(a) Illustrates the expected two-dimensional TPSC function $W(\omega_0+d',\omega_0+d)$ for partially coherent quantum pulses from a type-0 PDC (left). We plot the spectrum (green dashed line) and coherence profile (red solid curve) together to highlight that coherence width is much narrower than the spectral bandwidth (right). (b) Measured and theoretical spectral coherence profiles $|W(\omega_0,\omega_0+d)|$ alongside the expected spectrum $S(\omega_0-d)$. The maximum of each profile is normalized to 1. (c) Measured and theoretical Arg$[W(\omega_0,\omega_0+d)]$ profiles.}\label{fig5}
\end{figure}

To demonstrate the suitability of our scheme for quantum light, we apply it to a type-0 integrated parametric down-conversion (PDC) source. The experimental details of the source are provided in the Appendix Sec.~\ref{exp}. This source exhibits strong spectral-temporal correlations between the generated photon pairs, while each photon has large spectral and temporal widths. As a consequence, the quantum pulses generated by PDC have spectral and temporal bandwidths that are much broader than their corresponding coherence widths, resulting in partially coherent pulses. To further illustrate this partial coherence, the expected two-dimensional TPSC function is shown in Fig.~\ref{fig5}(a), which clearly reveals that the spectrum is significantly broader than the coherence profile. Additional details on the TF characteristics of the type-0 PDC process are provided in the supplementary material.

Figures~\ref{fig5}(c) and (d) show the measured $|W(\omega_0, \omega_0+d)|$ and Arg$[W(\omega_0, \omega_0+d)]$ alongside their theoretical predictions. {In Fig.~\ref{fig5}(c), the measured $W(\omega_0, \omega_0+d)$ is plotted together with the expected spectrum $S(\omega_0-d)$ (dashed green line), to highlight that the measured coherence profile is significantly narrower than the expected spectrum, demonstrating the low spectral coherence of the PDC pulses.} A quantitative comparison between the theoretical and experimental profiles yields a similarity $S \approx 98\%$. This demonstrates the effectiveness of our scheme for quantum pulses with low spectral coherence. {The data accumulation time of 30 minutes yields} high signal to noise ratio, {such that} the statistical errors in the reconstructed profiles are almost negligible. {The} flexibility of this scheme for characterizing pulses {is showcased} with varying degrees of coherence in the supplementary section. 


%
%
%
%
\begin{figure*}[t!]
	\centering
	\includegraphics[width=1\linewidth]{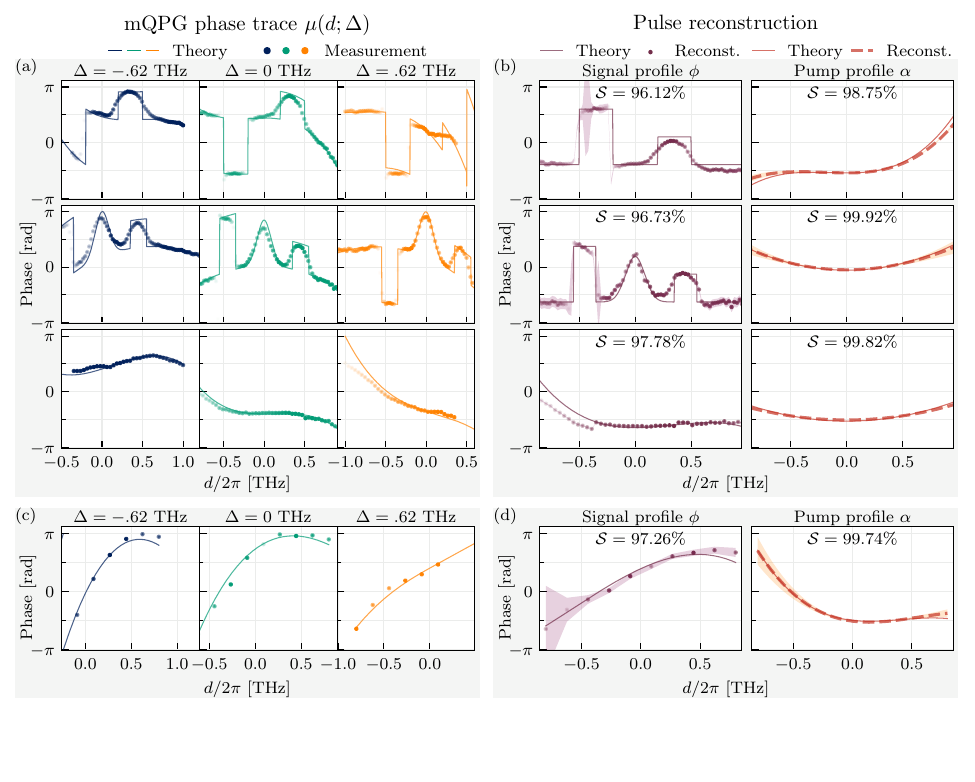}
	\caption{(a) Theoretical and measured phase traces, $\mu(d; \Delta)$, for $\Delta = -0.62$, $0$, and $0.62$ THz, corresponding to different spectral phase profiles of single-photon-level input pulses and bright pump pulses. (b) Reconstructed spectral phase profiles of the input and pump pulses for various shapes, shown alongside theoretical predictions. Shaded regions in the reconstructed profiles indicate statistical errors. (c) Theoretical and measured mQPG phase traces, $\mu(d; \Delta)$, for $\Delta = -0.62$, $0$, and $0.62$ THz, corresponding to low-light-level input and pump pulses. (d) Reconstructed spectral phase profiles of the pump and input pulses for different shapes, shown with theoretical predictions. Shaded regions in the reconstructed profiles indicate statistical errors.}\label{fig6}
\end{figure*}
%




%
%
%
%
%
	%
	%
	%
\subsection{Simultaneous characterization of spectral phase profiles of input and pump pulses}
{Finally, to showcase the capability of this scheme for characterizing single-photon level pulses without prior knowledge of pump pulse, in the experimental setup in Fig.~\ref{fig3} we replace the QPG with an mQPG.} {The perfectly coherent input pulses from the OPO are attenuated to a mean photon number of 1.0 per pulse, while the Ti:Sapphire laser pumps the mQPG. Data accumulation takes around 30 minutes.
See the Appendix Sec.~\ref{exp} for further details.

Figure~\ref{fig6}(a) shows the measured spectral phase profiles $\mu(d;\Delta)$ for different shearing values $\Delta$, corresponding to different input and pump phase shapes. We find a good agreement with the expected profiles. These measured profiles are then processed using the algorithm illustrated in Fig.~\ref{fig2} to retrieve the input and pump phase profiles. Figure~\ref{fig6}(b) shows the retrieved input and pump phase profiles and compares them with their corresponding theoretical predictions, showing a similarity above 96\%, even for complicated input phase structures. These results mark a major step toward self-referenced spectral phase characterization at the single-photon level, eliminating the limitation of requiring a known pump or reference pulse in existing phase characterization techniques. Moreover, the scheme remains applicable to single-photon pulses or lower fluxes with increased accumulation time.  
%

So far, we have operated with pump pulses of 62 pJ per pulse. To showcase the feasibility of our approach in extreme low-light conditions, we attenuate both the input and pump pulses to 1.25 fJ per pulse (~$10^4$ photons per pulse). As expected, the reduced photon flux, or pulse energy, increases the measurement time to approximately 60 minutes.
Figure~\ref{fig6}(c) presents the measured phase profiles for different shearing alongside their theoretical predictions. Figure~\ref{fig6}(d) shows the reconstructed input and pump phase profiles, indicating excellent agreement with theoretical predictions, with similarity $S$ above 97\%.  This demonstration highlights the applicability of our method for scenarios where simultaneous spectral phase characterization of two low-energy pulses at different wavelengths is required, such as in quantum metrology applications.

	
	%
\section{Conclusion and Discussion}
In conclusion, we have developed FIREFLY, a method for characterizing single-photon-level ultrafast pulses by measuring their TPSC functions using a QPG. This approach enables high-quality characterization of the spectral amplitude and phase profiles for spectrally coherent pulses. We have also {applied our scheme to quantum light} by {analysing} the spectral coherence profile of partially coherent pulses generated from a mutimode type-0 PDC process. Importantly, we have {then} eliminated the requirement for a spectrally known pump or reference pulse by extending FIREFLY to an mQPG. As proof of principle, we have demonstrated the simultaneous characterization of the spectral phase profiles for both single-photon input and bright pump pulses with complex spectral shapes. {Remarkably, our approach even works} {for both pulses} at the low-light-level.

Unlike most existing schemes, our method can characterize {complex phase structures and coherence profiles without a requiring reconstruction algorithm or setup reconfiguration. Additionally, the use of the mQPG to characterize single-photon-level pulses without a known reference pulse enhances experimental resource efficiency and ensures compatibility with the integrated optics platform. These features make our scheme readily suitable for quantum communication and metrology protocols that require fast characterization of single-photon pulses, as well as for deep-space communication in high-loss and turbulent environments.

Further improvements, such as adopting periodic poling modulation techniques \cite{chou1999optlet} in the mQPG fabrication and replacing the time-of-flight (TOF) spectrograph with a spectral-filtering-based detection scheme, can enhance efficiency by nearly 1000 fold. These improvements enable single-photon pulse characterization within seconds.

In our experiment, the QPG and the mQPG facilitate the characterization of single-photon-level telecom pulses through detection of the visible output photons, thereby eliminating the need for single-photon sensitive telecom detectors. This approach can be extended to input pulses of any wavelength \cite{roeder2024njp}, facilitating pulse characterization in spectral regions where single-photon-sensitive detectors are resource demanding. While our current FIREFLY implementation is optimal for characterizing input pulses with temporal widths ranging from 100 fs to 20 ps, this range can be extended by tailoring the dispersion properties of the waveguide \cite{roeder2024njp}. Furthermore, advances in thin-film lithium niobate technology \cite{zhu2021integrated} offer greater flexibility in modulating dispersion properties with tailored waveguide geometries, which can enable (m)QPGs compatible with sub-femtosecond input pulses across any spectral range. These developments can make our scheme valuable for diverse applications in quantum information science and beyond, including ultrafast optics and spectroscopy.

\section{Acknowledgements}
	We acknowledge the European Union’s Horizon Europe research and innovation programme under grant agreement No 899587 (STORMYTUNE) for the funding.
\section{Data Availability}
All data required for evaluating our conclusions are present in the paper and supplementary material. Additional data related to this paper may be requested from the corresponding author	
	
\bibliographystyle{apsrev4-2}
\bibliography{ref}

\begin{thebibliography}{50}%
\makeatletter
\providecommand \@ifxundefined [1]{%
 \@ifx{#1\undefined}
}%
\providecommand \@ifnum [1]{%
 \ifnum #1\expandafter \@firstoftwo
 \else \expandafter \@secondoftwo
 \fi
}%
\providecommand \@ifx [1]{%
 \ifx #1\expandafter \@firstoftwo
 \else \expandafter \@secondoftwo
 \fi
}%
\providecommand \natexlab [1]{#1}%
\providecommand \enquote  [1]{``#1''}%
\providecommand \bibnamefont  [1]{#1}%
\providecommand \bibfnamefont [1]{#1}%
\providecommand \citenamefont [1]{#1}%
\providecommand \href@noop [0]{\@secondoftwo}%
\providecommand \href [0]{\begingroup \@sanitize@url \@href}%
\providecommand \@href[1]{\@@startlink{#1}\@@href}%
\providecommand \@@href[1]{\endgroup#1\@@endlink}%
\providecommand \@sanitize@url [0]{\catcode `\\12\catcode `\$12\catcode `\&12\catcode `\#12\catcode `\^12\catcode `\_12\catcode `\%12\relax}%
\providecommand \@@startlink[1]{}%
\providecommand \@@endlink[0]{}%
\providecommand \url  [0]{\begingroup\@sanitize@url \@url }%
\providecommand \@url [1]{\endgroup\@href {#1}{\urlprefix }}%
\providecommand \urlprefix  [0]{URL }%
\providecommand \Eprint [0]{\href }%
\providecommand \doibase [0]{https://doi.org/}%
\providecommand \selectlanguage [0]{\@gobble}%
\providecommand \bibinfo  [0]{\@secondoftwo}%
\providecommand \bibfield  [0]{\@secondoftwo}%
\providecommand \translation [1]{[#1]}%
\providecommand \BibitemOpen [0]{}%
\providecommand \bibitemStop [0]{}%
\providecommand \bibitemNoStop [0]{.\EOS\space}%
\providecommand \EOS [0]{\spacefactor3000\relax}%
\providecommand \BibitemShut  [1]{\csname bibitem#1\endcsname}%
\let\auto@bib@innerbib\@empty
\bibitem [{\citenamefont {O'brien}\ \emph {et~al.}(2009)\citenamefont {O'brien}, \citenamefont {Furusawa},\ and\ \citenamefont {Vu{\v{c}}kovi{\'c}}}]{o2009photonic}%
  \BibitemOpen
  \bibfield  {author} {\bibinfo {author} {\bibfnamefont {J.~L.}\ \bibnamefont {O'brien}}, \bibinfo {author} {\bibfnamefont {A.}~\bibnamefont {Furusawa}},\ and\ \bibinfo {author} {\bibfnamefont {J.}~\bibnamefont {Vu{\v{c}}kovi{\'c}}},\ }\href@noop {} {\bibfield  {journal} {\bibinfo  {journal} {Nature Photonics}\ }\textbf {\bibinfo {volume} {3}},\ \bibinfo {pages} {687} (\bibinfo {year} {2009})}\BibitemShut {NoStop}%
\bibitem [{\citenamefont {Flamini}\ \emph {et~al.}(2018)\citenamefont {Flamini}, \citenamefont {Spagnolo},\ and\ \citenamefont {Sciarrino}}]{flamini2018photonic}%
  \BibitemOpen
  \bibfield  {author} {\bibinfo {author} {\bibfnamefont {F.}~\bibnamefont {Flamini}}, \bibinfo {author} {\bibfnamefont {N.}~\bibnamefont {Spagnolo}},\ and\ \bibinfo {author} {\bibfnamefont {F.}~\bibnamefont {Sciarrino}},\ }\href@noop {} {\bibfield  {journal} {\bibinfo  {journal} {Reports on Progress in Physics}\ }\textbf {\bibinfo {volume} {82}},\ \bibinfo {pages} {016001} (\bibinfo {year} {2018})}\BibitemShut {NoStop}%
\bibitem [{\citenamefont {Pelucchi}\ \emph {et~al.}(2022)\citenamefont {Pelucchi}, \citenamefont {Fagas}, \citenamefont {Aharonovich}, \citenamefont {Englund}, \citenamefont {Figueroa}, \citenamefont {Gong}, \citenamefont {Hannes}, \citenamefont {Liu}, \citenamefont {Lu}, \citenamefont {Matsuda} \emph {et~al.}}]{pelucchi2022potential}%
  \BibitemOpen
  \bibfield  {author} {\bibinfo {author} {\bibfnamefont {E.}~\bibnamefont {Pelucchi}}, \bibinfo {author} {\bibfnamefont {G.}~\bibnamefont {Fagas}}, \bibinfo {author} {\bibfnamefont {I.}~\bibnamefont {Aharonovich}}, \bibinfo {author} {\bibfnamefont {D.}~\bibnamefont {Englund}}, \bibinfo {author} {\bibfnamefont {E.}~\bibnamefont {Figueroa}}, \bibinfo {author} {\bibfnamefont {Q.}~\bibnamefont {Gong}}, \bibinfo {author} {\bibfnamefont {H.}~\bibnamefont {Hannes}}, \bibinfo {author} {\bibfnamefont {J.}~\bibnamefont {Liu}}, \bibinfo {author} {\bibfnamefont {C.-Y.}\ \bibnamefont {Lu}}, \bibinfo {author} {\bibfnamefont {N.}~\bibnamefont {Matsuda}}, \emph {et~al.},\ }\href@noop {} {\bibfield  {journal} {\bibinfo  {journal} {Nature Reviews Physics}\ }\textbf {\bibinfo {volume} {4}},\ \bibinfo {pages} {194} (\bibinfo {year} {2022})}\BibitemShut {NoStop}%
\bibitem [{\citenamefont {Brecht}\ \emph {et~al.}(2015)\citenamefont {Brecht}, \citenamefont {Reddy}, \citenamefont {Silberhorn},\ and\ \citenamefont {Raymer}}]{brechtprx2015}%
  \BibitemOpen
  \bibfield  {author} {\bibinfo {author} {\bibfnamefont {B.}~\bibnamefont {Brecht}}, \bibinfo {author} {\bibfnamefont {D.~V.}\ \bibnamefont {Reddy}}, \bibinfo {author} {\bibfnamefont {C.}~\bibnamefont {Silberhorn}},\ and\ \bibinfo {author} {\bibfnamefont {M.~G.}\ \bibnamefont {Raymer}},\ }\href {https://doi.org/10.1103/PhysRevX.5.041017} {\bibfield  {journal} {\bibinfo  {journal} {Phys. Rev. X}\ }\textbf {\bibinfo {volume} {5}},\ \bibinfo {pages} {041017} (\bibinfo {year} {2015})}\BibitemShut {NoStop}%
\bibitem [{\citenamefont {Ansari}\ \emph {et~al.}(2018{\natexlab{a}})\citenamefont {Ansari}, \citenamefont {Donohue}, \citenamefont {Brecht},\ and\ \citenamefont {Silberhorn}}]{ansari2018optica}%
  \BibitemOpen
  \bibfield  {author} {\bibinfo {author} {\bibfnamefont {V.}~\bibnamefont {Ansari}}, \bibinfo {author} {\bibfnamefont {J.~M.}\ \bibnamefont {Donohue}}, \bibinfo {author} {\bibfnamefont {B.}~\bibnamefont {Brecht}},\ and\ \bibinfo {author} {\bibfnamefont {C.}~\bibnamefont {Silberhorn}},\ }\href@noop {} {\bibfield  {journal} {\bibinfo  {journal} {Optica}\ }\textbf {\bibinfo {volume} {5}},\ \bibinfo {pages} {534} (\bibinfo {year} {2018}{\natexlab{a}})}\BibitemShut {NoStop}%
\bibitem [{\citenamefont {Karpi{\'n}ski}\ \emph {et~al.}(2021)\citenamefont {Karpi{\'n}ski}, \citenamefont {Davis}, \citenamefont {So{\'s}nicki}, \citenamefont {Thiel},\ and\ \citenamefont {Smith}}]{karpinski2021control}%
  \BibitemOpen
  \bibfield  {author} {\bibinfo {author} {\bibfnamefont {M.}~\bibnamefont {Karpi{\'n}ski}}, \bibinfo {author} {\bibfnamefont {A.~O.}\ \bibnamefont {Davis}}, \bibinfo {author} {\bibfnamefont {F.}~\bibnamefont {So{\'s}nicki}}, \bibinfo {author} {\bibfnamefont {V.}~\bibnamefont {Thiel}},\ and\ \bibinfo {author} {\bibfnamefont {B.~J.}\ \bibnamefont {Smith}},\ }\href@noop {} {\bibfield  {journal} {\bibinfo  {journal} {Advanced Quantum Technologies}\ }\textbf {\bibinfo {volume} {4}},\ \bibinfo {pages} {2000150} (\bibinfo {year} {2021})}\BibitemShut {NoStop}%
\bibitem [{\citenamefont {Humphreys}\ \emph {et~al.}(2013)\citenamefont {Humphreys}, \citenamefont {Metcalf}, \citenamefont {Spring}, \citenamefont {Moore}, \citenamefont {Jin}, \citenamefont {Barbieri}, \citenamefont {Kolthammer},\ and\ \citenamefont {Walmsley}}]{humphreys2013linear}%
  \BibitemOpen
  \bibfield  {author} {\bibinfo {author} {\bibfnamefont {P.~C.}\ \bibnamefont {Humphreys}}, \bibinfo {author} {\bibfnamefont {B.~J.}\ \bibnamefont {Metcalf}}, \bibinfo {author} {\bibfnamefont {J.~B.}\ \bibnamefont {Spring}}, \bibinfo {author} {\bibfnamefont {M.}~\bibnamefont {Moore}}, \bibinfo {author} {\bibfnamefont {X.-M.}\ \bibnamefont {Jin}}, \bibinfo {author} {\bibfnamefont {M.}~\bibnamefont {Barbieri}}, \bibinfo {author} {\bibfnamefont {W.~S.}\ \bibnamefont {Kolthammer}},\ and\ \bibinfo {author} {\bibfnamefont {I.~A.}\ \bibnamefont {Walmsley}},\ }\href@noop {} {\bibfield  {journal} {\bibinfo  {journal} {Physical review letters}\ }\textbf {\bibinfo {volume} {111}},\ \bibinfo {pages} {150501} (\bibinfo {year} {2013})}\BibitemShut {NoStop}%
\bibitem [{\citenamefont {Menicucci}\ \emph {et~al.}(2008)\citenamefont {Menicucci}, \citenamefont {Flammia},\ and\ \citenamefont {Pfister}}]{menicucci2008prl}%
  \BibitemOpen
  \bibfield  {author} {\bibinfo {author} {\bibfnamefont {N.~C.}\ \bibnamefont {Menicucci}}, \bibinfo {author} {\bibfnamefont {S.~T.}\ \bibnamefont {Flammia}},\ and\ \bibinfo {author} {\bibfnamefont {O.}~\bibnamefont {Pfister}},\ }\href@noop {} {\bibfield  {journal} {\bibinfo  {journal} {Physical review letters}\ }\textbf {\bibinfo {volume} {101}},\ \bibinfo {pages} {130501} (\bibinfo {year} {2008})}\BibitemShut {NoStop}%
\bibitem [{\citenamefont {Bouchard}\ \emph {et~al.}(2022)\citenamefont {Bouchard}, \citenamefont {England}, \citenamefont {Bustard}, \citenamefont {Heshami},\ and\ \citenamefont {Sussman}}]{bouchard2022prx}%
  \BibitemOpen
  \bibfield  {author} {\bibinfo {author} {\bibfnamefont {F.}~\bibnamefont {Bouchard}}, \bibinfo {author} {\bibfnamefont {D.}~\bibnamefont {England}}, \bibinfo {author} {\bibfnamefont {P.~J.}\ \bibnamefont {Bustard}}, \bibinfo {author} {\bibfnamefont {K.}~\bibnamefont {Heshami}},\ and\ \bibinfo {author} {\bibfnamefont {B.}~\bibnamefont {Sussman}},\ }\href@noop {} {\bibfield  {journal} {\bibinfo  {journal} {PRX Quantum}\ }\textbf {\bibinfo {volume} {3}},\ \bibinfo {pages} {010332} (\bibinfo {year} {2022})}\BibitemShut {NoStop}%
\bibitem [{\citenamefont {Bouchard}\ \emph {et~al.}(2021)\citenamefont {Bouchard}, \citenamefont {England}, \citenamefont {Bustard}, \citenamefont {Fenwick}, \citenamefont {Karimi}, \citenamefont {Heshami},\ and\ \citenamefont {Sussman}}]{bouchard2021achieving}%
  \BibitemOpen
  \bibfield  {author} {\bibinfo {author} {\bibfnamefont {F.}~\bibnamefont {Bouchard}}, \bibinfo {author} {\bibfnamefont {D.}~\bibnamefont {England}}, \bibinfo {author} {\bibfnamefont {P.~J.}\ \bibnamefont {Bustard}}, \bibinfo {author} {\bibfnamefont {K.~L.}\ \bibnamefont {Fenwick}}, \bibinfo {author} {\bibfnamefont {E.}~\bibnamefont {Karimi}}, \bibinfo {author} {\bibfnamefont {K.}~\bibnamefont {Heshami}},\ and\ \bibinfo {author} {\bibfnamefont {B.}~\bibnamefont {Sussman}},\ }\href@noop {} {\bibfield  {journal} {\bibinfo  {journal} {Physical Review Applied}\ }\textbf {\bibinfo {volume} {15}},\ \bibinfo {pages} {024027} (\bibinfo {year} {2021})}\BibitemShut {NoStop}%
\bibitem [{\citenamefont {Mukamel}\ \emph {et~al.}(2020)\citenamefont {Mukamel}, \citenamefont {Freyberger}, \citenamefont {Schleich}, \citenamefont {Bellini}, \citenamefont {Zavatta}, \citenamefont {Leuchs}, \citenamefont {Silberhorn}, \citenamefont {Boyd}, \citenamefont {S{\'a}nchez-Soto}, \citenamefont {Stefanov} \emph {et~al.}}]{mukamel2020JOPB}%
  \BibitemOpen
  \bibfield  {author} {\bibinfo {author} {\bibfnamefont {S.}~\bibnamefont {Mukamel}}, \bibinfo {author} {\bibfnamefont {M.}~\bibnamefont {Freyberger}}, \bibinfo {author} {\bibfnamefont {W.}~\bibnamefont {Schleich}}, \bibinfo {author} {\bibfnamefont {M.}~\bibnamefont {Bellini}}, \bibinfo {author} {\bibfnamefont {A.}~\bibnamefont {Zavatta}}, \bibinfo {author} {\bibfnamefont {G.}~\bibnamefont {Leuchs}}, \bibinfo {author} {\bibfnamefont {C.}~\bibnamefont {Silberhorn}}, \bibinfo {author} {\bibfnamefont {R.~W.}\ \bibnamefont {Boyd}}, \bibinfo {author} {\bibfnamefont {L.~L.}\ \bibnamefont {S{\'a}nchez-Soto}}, \bibinfo {author} {\bibfnamefont {A.}~\bibnamefont {Stefanov}}, \emph {et~al.},\ }\href@noop {} {\bibfield  {journal} {\bibinfo  {journal} {Journal of physics B: Atomic, molecular and optical physics}\ }\textbf {\bibinfo {volume} {53}},\ \bibinfo {pages} {072002} (\bibinfo {year} {2020})}\BibitemShut {NoStop}%
\bibitem [{\citenamefont {Donohue}\ \emph {et~al.}(2018)\citenamefont {Donohue}, \citenamefont {Ansari}, \citenamefont {{\v{R}}eh{\'a}{\v{c}}ek}, \citenamefont {Hradil}, \citenamefont {Stoklasa}, \citenamefont {Pa{\'u}r}, \citenamefont {S{\'a}nchez-Soto},\ and\ \citenamefont {Silberhorn}}]{donohue2018prl}%
  \BibitemOpen
  \bibfield  {author} {\bibinfo {author} {\bibfnamefont {J.~M.}\ \bibnamefont {Donohue}}, \bibinfo {author} {\bibfnamefont {V.}~\bibnamefont {Ansari}}, \bibinfo {author} {\bibfnamefont {J.}~\bibnamefont {{\v{R}}eh{\'a}{\v{c}}ek}}, \bibinfo {author} {\bibfnamefont {Z.}~\bibnamefont {Hradil}}, \bibinfo {author} {\bibfnamefont {B.}~\bibnamefont {Stoklasa}}, \bibinfo {author} {\bibfnamefont {M.}~\bibnamefont {Pa{\'u}r}}, \bibinfo {author} {\bibfnamefont {L.~L.}\ \bibnamefont {S{\'a}nchez-Soto}},\ and\ \bibinfo {author} {\bibfnamefont {C.}~\bibnamefont {Silberhorn}},\ }\href@noop {} {\bibfield  {journal} {\bibinfo  {journal} {Physical review letters}\ }\textbf {\bibinfo {volume} {121}},\ \bibinfo {pages} {090501} (\bibinfo {year} {2018})}\BibitemShut {NoStop}%
\bibitem [{\citenamefont {Giovannetti}\ \emph {et~al.}(2011)\citenamefont {Giovannetti}, \citenamefont {Lloyd},\ and\ \citenamefont {Maccone}}]{giovannetti2011advances}%
  \BibitemOpen
  \bibfield  {author} {\bibinfo {author} {\bibfnamefont {V.}~\bibnamefont {Giovannetti}}, \bibinfo {author} {\bibfnamefont {S.}~\bibnamefont {Lloyd}},\ and\ \bibinfo {author} {\bibfnamefont {L.}~\bibnamefont {Maccone}},\ }\href@noop {} {\bibfield  {journal} {\bibinfo  {journal} {Nature photonics}\ }\textbf {\bibinfo {volume} {5}},\ \bibinfo {pages} {222} (\bibinfo {year} {2011})}\BibitemShut {NoStop}%
\bibitem [{\citenamefont {Barbieri}(2022)}]{barbieri2022prxquantum}%
  \BibitemOpen
  \bibfield  {author} {\bibinfo {author} {\bibfnamefont {M.}~\bibnamefont {Barbieri}},\ }\href@noop {} {\bibfield  {journal} {\bibinfo  {journal} {PRX Quantum}\ }\textbf {\bibinfo {volume} {3}},\ \bibinfo {pages} {010202} (\bibinfo {year} {2022})}\BibitemShut {NoStop}%
\bibitem [{\citenamefont {Trebino}\ and\ \citenamefont {Kane}(1993)}]{Trebino:93}%
  \BibitemOpen
  \bibfield  {author} {\bibinfo {author} {\bibfnamefont {R.}~\bibnamefont {Trebino}}\ and\ \bibinfo {author} {\bibfnamefont {D.~J.}\ \bibnamefont {Kane}},\ }\href {https://doi.org/10.1364/JOSAA.10.001101} {\bibfield  {journal} {\bibinfo  {journal} {J. Opt. Soc. Am. A}\ }\textbf {\bibinfo {volume} {10}},\ \bibinfo {pages} {1101} (\bibinfo {year} {1993})}\BibitemShut {NoStop}%
\bibitem [{\citenamefont {DeLong}\ \emph {et~al.}(1994)\citenamefont {DeLong}, \citenamefont {Trebino}, \citenamefont {Hunter},\ and\ \citenamefont {White}}]{delong1994josab}%
  \BibitemOpen
  \bibfield  {author} {\bibinfo {author} {\bibfnamefont {K.}~\bibnamefont {DeLong}}, \bibinfo {author} {\bibfnamefont {R.}~\bibnamefont {Trebino}}, \bibinfo {author} {\bibfnamefont {J.}~\bibnamefont {Hunter}},\ and\ \bibinfo {author} {\bibfnamefont {W.}~\bibnamefont {White}},\ }\href@noop {} {\bibfield  {journal} {\bibinfo  {journal} {JOSA B}\ }\textbf {\bibinfo {volume} {11}},\ \bibinfo {pages} {2206} (\bibinfo {year} {1994})}\BibitemShut {NoStop}%
\bibitem [{\citenamefont {Walmsley}\ and\ \citenamefont {Wong}(1996)}]{walmsley1996josab}%
  \BibitemOpen
  \bibfield  {author} {\bibinfo {author} {\bibfnamefont {I.~A.}\ \bibnamefont {Walmsley}}\ and\ \bibinfo {author} {\bibfnamefont {V.}~\bibnamefont {Wong}},\ }\href@noop {} {\bibfield  {journal} {\bibinfo  {journal} {JOSA B}\ }\textbf {\bibinfo {volume} {13}},\ \bibinfo {pages} {2453} (\bibinfo {year} {1996})}\BibitemShut {NoStop}%
\bibitem [{\citenamefont {Iaconis}\ and\ \citenamefont {Walmsley}(1998)}]{iaconis1998optlett}%
  \BibitemOpen
  \bibfield  {author} {\bibinfo {author} {\bibfnamefont {C.}~\bibnamefont {Iaconis}}\ and\ \bibinfo {author} {\bibfnamefont {I.~A.}\ \bibnamefont {Walmsley}},\ }\href@noop {} {\bibfield  {journal} {\bibinfo  {journal} {Optics letters}\ }\textbf {\bibinfo {volume} {23}},\ \bibinfo {pages} {792} (\bibinfo {year} {1998})}\BibitemShut {NoStop}%
\bibitem [{\citenamefont {Londero}\ \emph {et~al.}(2003)\citenamefont {Londero}, \citenamefont {Anderson}, \citenamefont {Radzewicz}, \citenamefont {Iaconis},\ and\ \citenamefont {Walmsley}}]{londero2003jmo}%
  \BibitemOpen
  \bibfield  {author} {\bibinfo {author} {\bibfnamefont {P.}~\bibnamefont {Londero}}, \bibinfo {author} {\bibfnamefont {M.~E.}\ \bibnamefont {Anderson}}, \bibinfo {author} {\bibfnamefont {C.}~\bibnamefont {Radzewicz}}, \bibinfo {author} {\bibfnamefont {C.}~\bibnamefont {Iaconis}},\ and\ \bibinfo {author} {\bibfnamefont {I.~A.}\ \bibnamefont {Walmsley}},\ }\href@noop {} {\bibfield  {journal} {\bibinfo  {journal} {Journal of Modern Optics}\ }\textbf {\bibinfo {volume} {50}},\ \bibinfo {pages} {179} (\bibinfo {year} {2003})}\BibitemShut {NoStop}%
\bibitem [{\citenamefont {Fittinghoff}\ \emph {et~al.}(1996)\citenamefont {Fittinghoff}, \citenamefont {Bowie}, \citenamefont {Sweetser}, \citenamefont {Jennings}, \citenamefont {Krumb{\"u}gel}, \citenamefont {DeLong}, \citenamefont {Trebino},\ and\ \citenamefont {Walmsley}}]{fittinghoff1996optlett}%
  \BibitemOpen
  \bibfield  {author} {\bibinfo {author} {\bibfnamefont {D.~N.}\ \bibnamefont {Fittinghoff}}, \bibinfo {author} {\bibfnamefont {J.~L.}\ \bibnamefont {Bowie}}, \bibinfo {author} {\bibfnamefont {J.~N.}\ \bibnamefont {Sweetser}}, \bibinfo {author} {\bibfnamefont {R.~T.}\ \bibnamefont {Jennings}}, \bibinfo {author} {\bibfnamefont {M.~A.}\ \bibnamefont {Krumb{\"u}gel}}, \bibinfo {author} {\bibfnamefont {K.~W.}\ \bibnamefont {DeLong}}, \bibinfo {author} {\bibfnamefont {R.}~\bibnamefont {Trebino}},\ and\ \bibinfo {author} {\bibfnamefont {I.~A.}\ \bibnamefont {Walmsley}},\ }\href@noop {} {\bibfield  {journal} {\bibinfo  {journal} {Optics letters}\ }\textbf {\bibinfo {volume} {21}},\ \bibinfo {pages} {884} (\bibinfo {year} {1996})}\BibitemShut {NoStop}%
\bibitem [{\citenamefont {Gallmann}\ \emph {et~al.}(2001)\citenamefont {Gallmann}, \citenamefont {Steinmeyer}, \citenamefont {Sutter}, \citenamefont {Rupp}, \citenamefont {Iaconis}, \citenamefont {Walmsley},\ and\ \citenamefont {Keller}}]{gallmann2001optlett}%
  \BibitemOpen
  \bibfield  {author} {\bibinfo {author} {\bibfnamefont {L.}~\bibnamefont {Gallmann}}, \bibinfo {author} {\bibfnamefont {G.}~\bibnamefont {Steinmeyer}}, \bibinfo {author} {\bibfnamefont {D.}~\bibnamefont {Sutter}}, \bibinfo {author} {\bibfnamefont {T.}~\bibnamefont {Rupp}}, \bibinfo {author} {\bibfnamefont {C.}~\bibnamefont {Iaconis}}, \bibinfo {author} {\bibfnamefont {I.}~\bibnamefont {Walmsley}},\ and\ \bibinfo {author} {\bibfnamefont {U.}~\bibnamefont {Keller}},\ }\href@noop {} {\bibfield  {journal} {\bibinfo  {journal} {Optics Letters}\ }\textbf {\bibinfo {volume} {26}},\ \bibinfo {pages} {96} (\bibinfo {year} {2001})}\BibitemShut {NoStop}%
\bibitem [{\citenamefont {Kacprowicz}\ \emph {et~al.}(2008)\citenamefont {Kacprowicz}, \citenamefont {Wasilewski},\ and\ \citenamefont {Banaszek}}]{kacprowicz2008complete}%
  \BibitemOpen
  \bibfield  {author} {\bibinfo {author} {\bibfnamefont {M.}~\bibnamefont {Kacprowicz}}, \bibinfo {author} {\bibfnamefont {W.}~\bibnamefont {Wasilewski}},\ and\ \bibinfo {author} {\bibfnamefont {K.}~\bibnamefont {Banaszek}},\ }\href@noop {} {\bibfield  {journal} {\bibinfo  {journal} {Applied Physics B}\ }\textbf {\bibinfo {volume} {91}},\ \bibinfo {pages} {283} (\bibinfo {year} {2008})}\BibitemShut {NoStop}%
\bibitem [{\citenamefont {Reid}\ \emph {et~al.}(2000)\citenamefont {Reid}, \citenamefont {Loza-Alvarez}, \citenamefont {Brown}, \citenamefont {Beddard},\ and\ \citenamefont {Sibbett}}]{reid2000optlett}%
  \BibitemOpen
  \bibfield  {author} {\bibinfo {author} {\bibfnamefont {D.}~\bibnamefont {Reid}}, \bibinfo {author} {\bibfnamefont {P.}~\bibnamefont {Loza-Alvarez}}, \bibinfo {author} {\bibfnamefont {C.}~\bibnamefont {Brown}}, \bibinfo {author} {\bibfnamefont {T.}~\bibnamefont {Beddard}},\ and\ \bibinfo {author} {\bibfnamefont {W.}~\bibnamefont {Sibbett}},\ }\href@noop {} {\bibfield  {journal} {\bibinfo  {journal} {Optics letters}\ }\textbf {\bibinfo {volume} {25}},\ \bibinfo {pages} {1478} (\bibinfo {year} {2000})}\BibitemShut {NoStop}%
\bibitem [{\citenamefont {Stibenz}\ and\ \citenamefont {Steinmeyer}(2005)}]{stibenz2005optlett}%
  \BibitemOpen
  \bibfield  {author} {\bibinfo {author} {\bibfnamefont {G.}~\bibnamefont {Stibenz}}\ and\ \bibinfo {author} {\bibfnamefont {G.}~\bibnamefont {Steinmeyer}},\ }\href@noop {} {\bibfield  {journal} {\bibinfo  {journal} {Optics express}\ }\textbf {\bibinfo {volume} {13}},\ \bibinfo {pages} {2617} (\bibinfo {year} {2005})}\BibitemShut {NoStop}%
\bibitem [{\citenamefont {Pasquazi}\ \emph {et~al.}(2011)\citenamefont {Pasquazi}, \citenamefont {Peccianti}, \citenamefont {Park}, \citenamefont {Little}, \citenamefont {Chu}, \citenamefont {Morandotti}, \citenamefont {Aza{\~n}a},\ and\ \citenamefont {Moss}}]{pasquazi2011sub}%
  \BibitemOpen
  \bibfield  {author} {\bibinfo {author} {\bibfnamefont {A.}~\bibnamefont {Pasquazi}}, \bibinfo {author} {\bibfnamefont {M.}~\bibnamefont {Peccianti}}, \bibinfo {author} {\bibfnamefont {Y.}~\bibnamefont {Park}}, \bibinfo {author} {\bibfnamefont {B.~E.}\ \bibnamefont {Little}}, \bibinfo {author} {\bibfnamefont {S.~T.}\ \bibnamefont {Chu}}, \bibinfo {author} {\bibfnamefont {R.}~\bibnamefont {Morandotti}}, \bibinfo {author} {\bibfnamefont {J.}~\bibnamefont {Aza{\~n}a}},\ and\ \bibinfo {author} {\bibfnamefont {D.~J.}\ \bibnamefont {Moss}},\ }\href@noop {} {\bibfield  {journal} {\bibinfo  {journal} {Nature Photonics}\ }\textbf {\bibinfo {volume} {5}},\ \bibinfo {pages} {618} (\bibinfo {year} {2011})}\BibitemShut {NoStop}%
\bibitem [{\citenamefont {Bourassin-Bouchet}\ and\ \citenamefont {Couprie}(2015)}]{bourassin2015partially}%
  \BibitemOpen
  \bibfield  {author} {\bibinfo {author} {\bibfnamefont {C.}~\bibnamefont {Bourassin-Bouchet}}\ and\ \bibinfo {author} {\bibfnamefont {M.-E.}\ \bibnamefont {Couprie}},\ }\href@noop {} {\bibfield  {journal} {\bibinfo  {journal} {Nature communications}\ }\textbf {\bibinfo {volume} {6}},\ \bibinfo {pages} {6465} (\bibinfo {year} {2015})}\BibitemShut {NoStop}%
\bibitem [{\citenamefont {Davis}\ \emph {et~al.}(2018{\natexlab{a}})\citenamefont {Davis}, \citenamefont {Thiel}, \citenamefont {Karpi{\'n}ski},\ and\ \citenamefont {Smith}}]{davis2018pra}%
  \BibitemOpen
  \bibfield  {author} {\bibinfo {author} {\bibfnamefont {A.~O.}\ \bibnamefont {Davis}}, \bibinfo {author} {\bibfnamefont {V.}~\bibnamefont {Thiel}}, \bibinfo {author} {\bibfnamefont {M.}~\bibnamefont {Karpi{\'n}ski}},\ and\ \bibinfo {author} {\bibfnamefont {B.~J.}\ \bibnamefont {Smith}},\ }\href@noop {} {\bibfield  {journal} {\bibinfo  {journal} {Physical Review A}\ }\textbf {\bibinfo {volume} {98}},\ \bibinfo {pages} {023840} (\bibinfo {year} {2018}{\natexlab{a}})}\BibitemShut {NoStop}%
\bibitem [{\citenamefont {Davis}\ \emph {et~al.}(2018{\natexlab{b}})\citenamefont {Davis}, \citenamefont {Thiel}, \citenamefont {Karpi{\'n}ski},\ and\ \citenamefont {Smith}}]{davis2018prl}%
  \BibitemOpen
  \bibfield  {author} {\bibinfo {author} {\bibfnamefont {A.~O.}\ \bibnamefont {Davis}}, \bibinfo {author} {\bibfnamefont {V.}~\bibnamefont {Thiel}}, \bibinfo {author} {\bibfnamefont {M.}~\bibnamefont {Karpi{\'n}ski}},\ and\ \bibinfo {author} {\bibfnamefont {B.~J.}\ \bibnamefont {Smith}},\ }\href@noop {} {\bibfield  {journal} {\bibinfo  {journal} {Physical review letters}\ }\textbf {\bibinfo {volume} {121}},\ \bibinfo {pages} {083602} (\bibinfo {year} {2018}{\natexlab{b}})}\BibitemShut {NoStop}%
\bibitem [{\citenamefont {Golestani}\ \emph {et~al.}(2022)\citenamefont {Golestani}, \citenamefont {Davis}, \citenamefont {So{\'s}nicki}, \citenamefont {Miko{\l}ajczyk}, \citenamefont {Treps},\ and\ \citenamefont {Karpi{\'n}ski}}]{golestani2022prl}%
  \BibitemOpen
  \bibfield  {author} {\bibinfo {author} {\bibfnamefont {A.}~\bibnamefont {Golestani}}, \bibinfo {author} {\bibfnamefont {A.~O.}\ \bibnamefont {Davis}}, \bibinfo {author} {\bibfnamefont {F.}~\bibnamefont {So{\'s}nicki}}, \bibinfo {author} {\bibfnamefont {M.}~\bibnamefont {Miko{\l}ajczyk}}, \bibinfo {author} {\bibfnamefont {N.}~\bibnamefont {Treps}},\ and\ \bibinfo {author} {\bibfnamefont {M.}~\bibnamefont {Karpi{\'n}ski}},\ }\href@noop {} {\bibfield  {journal} {\bibinfo  {journal} {Physical Review Letters}\ }\textbf {\bibinfo {volume} {129}},\ \bibinfo {pages} {123605} (\bibinfo {year} {2022})}\BibitemShut {NoStop}%
\bibitem [{\citenamefont {Thiel}\ \emph {et~al.}(2020)\citenamefont {Thiel}, \citenamefont {Davis}, \citenamefont {Sun}, \citenamefont {D'Ornellas}, \citenamefont {Jin},\ and\ \citenamefont {Smith}}]{Thiel:20}%
  \BibitemOpen
  \bibfield  {author} {\bibinfo {author} {\bibfnamefont {V.}~\bibnamefont {Thiel}}, \bibinfo {author} {\bibfnamefont {A.~O.~C.}\ \bibnamefont {Davis}}, \bibinfo {author} {\bibfnamefont {K.}~\bibnamefont {Sun}}, \bibinfo {author} {\bibfnamefont {P.}~\bibnamefont {D'Ornellas}}, \bibinfo {author} {\bibfnamefont {X.-M.}\ \bibnamefont {Jin}},\ and\ \bibinfo {author} {\bibfnamefont {B.~J.}\ \bibnamefont {Smith}},\ }\href {https://doi.org/10.1364/OE.396960} {\bibfield  {journal} {\bibinfo  {journal} {Opt. Express}\ }\textbf {\bibinfo {volume} {28}},\ \bibinfo {pages} {19315} (\bibinfo {year} {2020})}\BibitemShut {NoStop}%
\bibitem [{\citenamefont {Lipka}\ and\ \citenamefont {Parniak}(2021)}]{lipka2021prl}%
  \BibitemOpen
  \bibfield  {author} {\bibinfo {author} {\bibfnamefont {M.}~\bibnamefont {Lipka}}\ and\ \bibinfo {author} {\bibfnamefont {M.}~\bibnamefont {Parniak}},\ }\href@noop {} {\bibfield  {journal} {\bibinfo  {journal} {Physical Review Letters}\ }\textbf {\bibinfo {volume} {127}},\ \bibinfo {pages} {163601} (\bibinfo {year} {2021})}\BibitemShut {NoStop}%
\bibitem [{\citenamefont {Thekkadath}\ \emph {et~al.}(2022)\citenamefont {Thekkadath}, \citenamefont {Bell}, \citenamefont {Patel}, \citenamefont {Kim},\ and\ \citenamefont {Walmsley}}]{thekkadath2022prl}%
  \BibitemOpen
  \bibfield  {author} {\bibinfo {author} {\bibfnamefont {G.}~\bibnamefont {Thekkadath}}, \bibinfo {author} {\bibfnamefont {B.}~\bibnamefont {Bell}}, \bibinfo {author} {\bibfnamefont {R.}~\bibnamefont {Patel}}, \bibinfo {author} {\bibfnamefont {M.}~\bibnamefont {Kim}},\ and\ \bibinfo {author} {\bibfnamefont {I.}~\bibnamefont {Walmsley}},\ }\href@noop {} {\bibfield  {journal} {\bibinfo  {journal} {Physical Review Letters}\ }\textbf {\bibinfo {volume} {128}},\ \bibinfo {pages} {023601} (\bibinfo {year} {2022})}\BibitemShut {NoStop}%
\bibitem [{\citenamefont {Qin}\ \emph {et~al.}(2015)\citenamefont {Qin}, \citenamefont {Prasad}, \citenamefont {Brannan}, \citenamefont {MacRae}, \citenamefont {Lezama},\ and\ \citenamefont {Lvovsky}}]{qin2015complete}%
  \BibitemOpen
  \bibfield  {author} {\bibinfo {author} {\bibfnamefont {Z.}~\bibnamefont {Qin}}, \bibinfo {author} {\bibfnamefont {A.~S.}\ \bibnamefont {Prasad}}, \bibinfo {author} {\bibfnamefont {T.}~\bibnamefont {Brannan}}, \bibinfo {author} {\bibfnamefont {A.}~\bibnamefont {MacRae}}, \bibinfo {author} {\bibfnamefont {A.}~\bibnamefont {Lezama}},\ and\ \bibinfo {author} {\bibfnamefont {A.}~\bibnamefont {Lvovsky}},\ }\href@noop {} {\bibfield  {journal} {\bibinfo  {journal} {Light: Science \& Applications}\ }\textbf {\bibinfo {volume} {4}},\ \bibinfo {pages} {e298} (\bibinfo {year} {2015})}\BibitemShut {NoStop}%
\bibitem [{\citenamefont {Wasilewski}\ \emph {et~al.}(2007)\citenamefont {Wasilewski}, \citenamefont {Kolenderski},\ and\ \citenamefont {Frankowski}}]{wasilewski2007prl}%
  \BibitemOpen
  \bibfield  {author} {\bibinfo {author} {\bibfnamefont {W.}~\bibnamefont {Wasilewski}}, \bibinfo {author} {\bibfnamefont {P.}~\bibnamefont {Kolenderski}},\ and\ \bibinfo {author} {\bibfnamefont {R.}~\bibnamefont {Frankowski}},\ }\href@noop {} {\bibfield  {journal} {\bibinfo  {journal} {Physical review letters}\ }\textbf {\bibinfo {volume} {99}},\ \bibinfo {pages} {123601} (\bibinfo {year} {2007})}\BibitemShut {NoStop}%
\bibitem [{\citenamefont {Bhattacharjee}\ \emph {et~al.}(2024)\citenamefont {Bhattacharjee}, \citenamefont {Serino}, \citenamefont {{\v{R}}eh{\'a}{\v{c}}ek}, \citenamefont {Hradil}, \citenamefont {Silberhorn},\ and\ \citenamefont {Brecht}}]{bhattacharjee2024pulse}%
  \BibitemOpen
  \bibfield  {author} {\bibinfo {author} {\bibfnamefont {A.}~\bibnamefont {Bhattacharjee}}, \bibinfo {author} {\bibfnamefont {L.}~\bibnamefont {Serino}}, \bibinfo {author} {\bibfnamefont {J.}~\bibnamefont {{\v{R}}eh{\'a}{\v{c}}ek}}, \bibinfo {author} {\bibfnamefont {Z.}~\bibnamefont {Hradil}}, \bibinfo {author} {\bibfnamefont {C.}~\bibnamefont {Silberhorn}},\ and\ \bibinfo {author} {\bibfnamefont {B.}~\bibnamefont {Brecht}},\ }\href@noop {} {\bibfield  {journal} {\bibinfo  {journal} {arXiv preprint arXiv:2408.12306}\ } (\bibinfo {year} {2024})}\BibitemShut {NoStop}%
\bibitem [{\citenamefont {Mandel}(1995)}]{mandel1995optical}%
  \BibitemOpen
  \bibfield  {author} {\bibinfo {author} {\bibfnamefont {L.}~\bibnamefont {Mandel}},\ }\href@noop {} {\emph {\bibinfo {title} {Optical Coherence and Quantum Optics}}}\ (\bibinfo  {publisher} {Cambridge University Press},\ \bibinfo {year} {1995})\BibitemShut {NoStop}%
\bibitem [{\citenamefont {Kulkarni}\ \emph {et~al.}(2017)\citenamefont {Kulkarni}, \citenamefont {Sahu}, \citenamefont {Maga{\~n}a-Loaiza}, \citenamefont {Boyd},\ and\ \citenamefont {Jha}}]{kulkarni2017single}%
  \BibitemOpen
  \bibfield  {author} {\bibinfo {author} {\bibfnamefont {G.}~\bibnamefont {Kulkarni}}, \bibinfo {author} {\bibfnamefont {R.}~\bibnamefont {Sahu}}, \bibinfo {author} {\bibfnamefont {O.~S.}\ \bibnamefont {Maga{\~n}a-Loaiza}}, \bibinfo {author} {\bibfnamefont {R.~W.}\ \bibnamefont {Boyd}},\ and\ \bibinfo {author} {\bibfnamefont {A.~K.}\ \bibnamefont {Jha}},\ }\href@noop {} {\bibfield  {journal} {\bibinfo  {journal} {Nature communications}\ }\textbf {\bibinfo {volume} {8}},\ \bibinfo {pages} {1054} (\bibinfo {year} {2017})}\BibitemShut {NoStop}%
\bibitem [{\citenamefont {Iaconis}\ and\ \citenamefont {Walmsley}(1996)}]{iaconis1996direct}%
  \BibitemOpen
  \bibfield  {author} {\bibinfo {author} {\bibfnamefont {C.}~\bibnamefont {Iaconis}}\ and\ \bibinfo {author} {\bibfnamefont {I.~A.}\ \bibnamefont {Walmsley}},\ }\href@noop {} {\bibfield  {journal} {\bibinfo  {journal} {Optics letters}\ }\textbf {\bibinfo {volume} {21}},\ \bibinfo {pages} {1783} (\bibinfo {year} {1996})}\BibitemShut {NoStop}%
\bibitem [{\citenamefont {Rezvani~Naraghi}\ \emph {et~al.}(2017)\citenamefont {Rezvani~Naraghi}, \citenamefont {Gemar}, \citenamefont {Batarseh}, \citenamefont {Beckus}, \citenamefont {Atia}, \citenamefont {Sukhov},\ and\ \citenamefont {Dogariu}}]{rezvani2017wide}%
  \BibitemOpen
  \bibfield  {author} {\bibinfo {author} {\bibfnamefont {R.}~\bibnamefont {Rezvani~Naraghi}}, \bibinfo {author} {\bibfnamefont {H.}~\bibnamefont {Gemar}}, \bibinfo {author} {\bibfnamefont {M.}~\bibnamefont {Batarseh}}, \bibinfo {author} {\bibfnamefont {A.}~\bibnamefont {Beckus}}, \bibinfo {author} {\bibfnamefont {G.}~\bibnamefont {Atia}}, \bibinfo {author} {\bibfnamefont {S.}~\bibnamefont {Sukhov}},\ and\ \bibinfo {author} {\bibfnamefont {A.}~\bibnamefont {Dogariu}},\ }\href@noop {} {\bibfield  {journal} {\bibinfo  {journal} {Optics letters}\ }\textbf {\bibinfo {volume} {42}},\ \bibinfo {pages} {4929} (\bibinfo {year} {2017})}\BibitemShut {NoStop}%
\bibitem [{\citenamefont {Bhattacharjee}\ \emph {et~al.}(2018)\citenamefont {Bhattacharjee}, \citenamefont {Aarav},\ and\ \citenamefont {Jha}}]{bhattacharjee2018apl}%
  \BibitemOpen
  \bibfield  {author} {\bibinfo {author} {\bibfnamefont {A.}~\bibnamefont {Bhattacharjee}}, \bibinfo {author} {\bibfnamefont {S.}~\bibnamefont {Aarav}},\ and\ \bibinfo {author} {\bibfnamefont {A.~K.}\ \bibnamefont {Jha}},\ }\href@noop {} {\bibfield  {journal} {\bibinfo  {journal} {Applied Physics Letters}\ }\textbf {\bibinfo {volume} {113}},\ \bibinfo {pages} {051102} (\bibinfo {year} {2018})}\BibitemShut {NoStop}%
\bibitem [{\citenamefont {Ansari}\ \emph {et~al.}(2017)\citenamefont {Ansari}, \citenamefont {Harder}, \citenamefont {Allgaier}, \citenamefont {Brecht},\ and\ \citenamefont {Silberhorn}}]{Ansaripra2017}%
  \BibitemOpen
  \bibfield  {author} {\bibinfo {author} {\bibfnamefont {V.}~\bibnamefont {Ansari}}, \bibinfo {author} {\bibfnamefont {G.}~\bibnamefont {Harder}}, \bibinfo {author} {\bibfnamefont {M.}~\bibnamefont {Allgaier}}, \bibinfo {author} {\bibfnamefont {B.}~\bibnamefont {Brecht}},\ and\ \bibinfo {author} {\bibfnamefont {C.}~\bibnamefont {Silberhorn}},\ }\href {https://doi.org/10.1103/PhysRevA.96.063817} {\bibfield  {journal} {\bibinfo  {journal} {Phys. Rev. A}\ }\textbf {\bibinfo {volume} {96}},\ \bibinfo {pages} {063817} (\bibinfo {year} {2017})}\BibitemShut {NoStop}%
\bibitem [{\citenamefont {Allgaier}\ \emph {et~al.}(2017)\citenamefont {Allgaier}, \citenamefont {Vigh}, \citenamefont {Ansari}, \citenamefont {Eigner}, \citenamefont {Quiring}, \citenamefont {Ricken}, \citenamefont {Brecht},\ and\ \citenamefont {Silberhorn}}]{allgaier2017QST}%
  \BibitemOpen
  \bibfield  {author} {\bibinfo {author} {\bibfnamefont {M.}~\bibnamefont {Allgaier}}, \bibinfo {author} {\bibfnamefont {G.}~\bibnamefont {Vigh}}, \bibinfo {author} {\bibfnamefont {V.}~\bibnamefont {Ansari}}, \bibinfo {author} {\bibfnamefont {C.}~\bibnamefont {Eigner}}, \bibinfo {author} {\bibfnamefont {V.}~\bibnamefont {Quiring}}, \bibinfo {author} {\bibfnamefont {R.}~\bibnamefont {Ricken}}, \bibinfo {author} {\bibfnamefont {B.}~\bibnamefont {Brecht}},\ and\ \bibinfo {author} {\bibfnamefont {C.}~\bibnamefont {Silberhorn}},\ }\href@noop {} {\bibfield  {journal} {\bibinfo  {journal} {Quantum Science and Technology}\ }\textbf {\bibinfo {volume} {2}},\ \bibinfo {pages} {034012} (\bibinfo {year} {2017})}\BibitemShut {NoStop}%
\bibitem [{\citenamefont {Ansari}\ \emph {et~al.}(2018{\natexlab{b}})\citenamefont {Ansari}, \citenamefont {Donohue}, \citenamefont {Allgaier}, \citenamefont {Sansoni}, \citenamefont {Brecht}, \citenamefont {Roslund}, \citenamefont {Treps}, \citenamefont {Harder},\ and\ \citenamefont {Silberhorn}}]{ansari2018prl}%
  \BibitemOpen
  \bibfield  {author} {\bibinfo {author} {\bibfnamefont {V.}~\bibnamefont {Ansari}}, \bibinfo {author} {\bibfnamefont {J.~M.}\ \bibnamefont {Donohue}}, \bibinfo {author} {\bibfnamefont {M.}~\bibnamefont {Allgaier}}, \bibinfo {author} {\bibfnamefont {L.}~\bibnamefont {Sansoni}}, \bibinfo {author} {\bibfnamefont {B.}~\bibnamefont {Brecht}}, \bibinfo {author} {\bibfnamefont {J.}~\bibnamefont {Roslund}}, \bibinfo {author} {\bibfnamefont {N.}~\bibnamefont {Treps}}, \bibinfo {author} {\bibfnamefont {G.}~\bibnamefont {Harder}},\ and\ \bibinfo {author} {\bibfnamefont {C.}~\bibnamefont {Silberhorn}},\ }\href@noop {} {\bibfield  {journal} {\bibinfo  {journal} {Physical review letters}\ }\textbf {\bibinfo {volume} {120}},\ \bibinfo {pages} {213601} (\bibinfo {year} {2018}{\natexlab{b}})}\BibitemShut {NoStop}%
\bibitem [{\citenamefont {Gil-Lopez}\ \emph {et~al.}(2021)\citenamefont {Gil-Lopez}, \citenamefont {Teo}, \citenamefont {De}, \citenamefont {Brecht}, \citenamefont {Jeong}, \citenamefont {Silberhorn},\ and\ \citenamefont {S{\'a}nchez-Soto}}]{gil2021optical}%
  \BibitemOpen
  \bibfield  {author} {\bibinfo {author} {\bibfnamefont {J.}~\bibnamefont {Gil-Lopez}}, \bibinfo {author} {\bibfnamefont {Y.~S.}\ \bibnamefont {Teo}}, \bibinfo {author} {\bibfnamefont {S.}~\bibnamefont {De}}, \bibinfo {author} {\bibfnamefont {B.}~\bibnamefont {Brecht}}, \bibinfo {author} {\bibfnamefont {H.}~\bibnamefont {Jeong}}, \bibinfo {author} {\bibfnamefont {C.}~\bibnamefont {Silberhorn}},\ and\ \bibinfo {author} {\bibfnamefont {L.~L.}\ \bibnamefont {S{\'a}nchez-Soto}},\ }\href@noop {} {\bibfield  {journal} {\bibinfo  {journal} {Optica}\ }\textbf {\bibinfo {volume} {8}},\ \bibinfo {pages} {1296} (\bibinfo {year} {2021})}\BibitemShut {NoStop}%
\bibitem [{\citenamefont {Ansari}\ \emph {et~al.}(2020)\citenamefont {Ansari}, \citenamefont {Donohue}, \citenamefont {Brecht},\ and\ \citenamefont {Silberhorn}}]{ansari2020optexp}%
  \BibitemOpen
  \bibfield  {author} {\bibinfo {author} {\bibfnamefont {V.}~\bibnamefont {Ansari}}, \bibinfo {author} {\bibfnamefont {J.~M.}\ \bibnamefont {Donohue}}, \bibinfo {author} {\bibfnamefont {B.}~\bibnamefont {Brecht}},\ and\ \bibinfo {author} {\bibfnamefont {C.}~\bibnamefont {Silberhorn}},\ }\href@noop {} {\bibfield  {journal} {\bibinfo  {journal} {Optics Express}\ }\textbf {\bibinfo {volume} {28}},\ \bibinfo {pages} {28295} (\bibinfo {year} {2020})}\BibitemShut {NoStop}%
\bibitem [{\citenamefont {Serino}\ \emph {et~al.}(2023)\citenamefont {Serino}, \citenamefont {Gil-Lopez}, \citenamefont {Stefszky}, \citenamefont {Ricken}, \citenamefont {Eigner}, \citenamefont {Brecht},\ and\ \citenamefont {Silberhorn}}]{serino2023prx}%
  \BibitemOpen
  \bibfield  {author} {\bibinfo {author} {\bibfnamefont {L.}~\bibnamefont {Serino}}, \bibinfo {author} {\bibfnamefont {J.}~\bibnamefont {Gil-Lopez}}, \bibinfo {author} {\bibfnamefont {M.}~\bibnamefont {Stefszky}}, \bibinfo {author} {\bibfnamefont {R.}~\bibnamefont {Ricken}}, \bibinfo {author} {\bibfnamefont {C.}~\bibnamefont {Eigner}}, \bibinfo {author} {\bibfnamefont {B.}~\bibnamefont {Brecht}},\ and\ \bibinfo {author} {\bibfnamefont {C.}~\bibnamefont {Silberhorn}},\ }\href@noop {} {\bibfield  {journal} {\bibinfo  {journal} {PRX quantum}\ }\textbf {\bibinfo {volume} {4}},\ \bibinfo {pages} {020306} (\bibinfo {year} {2023})}\BibitemShut {NoStop}%
\bibitem [{\citenamefont {Chou}\ \emph {et~al.}(1999)\citenamefont {Chou}, \citenamefont {Parameswaran}, \citenamefont {Fejer},\ and\ \citenamefont {Brener}}]{chou1999optlet}%
  \BibitemOpen
  \bibfield  {author} {\bibinfo {author} {\bibfnamefont {M.}~\bibnamefont {Chou}}, \bibinfo {author} {\bibfnamefont {K.}~\bibnamefont {Parameswaran}}, \bibinfo {author} {\bibfnamefont {M.~M.}\ \bibnamefont {Fejer}},\ and\ \bibinfo {author} {\bibfnamefont {I.}~\bibnamefont {Brener}},\ }\href@noop {} {\bibfield  {journal} {\bibinfo  {journal} {Optics letters}\ }\textbf {\bibinfo {volume} {24}},\ \bibinfo {pages} {1157} (\bibinfo {year} {1999})}\BibitemShut {NoStop}%
\bibitem [{\citenamefont {Roeder}\ \emph {et~al.}(2024)\citenamefont {Roeder}, \citenamefont {Gnanavel}, \citenamefont {Pollmann}, \citenamefont {Brecht}, \citenamefont {Stefszky}, \citenamefont {Padberg}, \citenamefont {Eigner}, \citenamefont {Silberhorn},\ and\ \citenamefont {Brecht}}]{roeder2024njp}%
  \BibitemOpen
  \bibfield  {author} {\bibinfo {author} {\bibfnamefont {F.}~\bibnamefont {Roeder}}, \bibinfo {author} {\bibfnamefont {A.}~\bibnamefont {Gnanavel}}, \bibinfo {author} {\bibfnamefont {R.}~\bibnamefont {Pollmann}}, \bibinfo {author} {\bibfnamefont {O.}~\bibnamefont {Brecht}}, \bibinfo {author} {\bibfnamefont {M.}~\bibnamefont {Stefszky}}, \bibinfo {author} {\bibfnamefont {L.}~\bibnamefont {Padberg}}, \bibinfo {author} {\bibfnamefont {C.}~\bibnamefont {Eigner}}, \bibinfo {author} {\bibfnamefont {C.}~\bibnamefont {Silberhorn}},\ and\ \bibinfo {author} {\bibfnamefont {B.}~\bibnamefont {Brecht}},\ }\href@noop {} {\bibfield  {journal} {\bibinfo  {journal} {New Journal of Physics}\ }\textbf {\bibinfo {volume} {26}},\ \bibinfo {pages} {123025} (\bibinfo {year} {2024})}\BibitemShut {NoStop}%
\bibitem [{\citenamefont {Zhu}\ \emph {et~al.}(2021)\citenamefont {Zhu}, \citenamefont {Shao}, \citenamefont {Yu}, \citenamefont {Cheng}, \citenamefont {Desiatov}, \citenamefont {Xin}, \citenamefont {Hu}, \citenamefont {Holzgrafe}, \citenamefont {Ghosh}, \citenamefont {Shams-Ansari} \emph {et~al.}}]{zhu2021integrated}%
  \BibitemOpen
  \bibfield  {author} {\bibinfo {author} {\bibfnamefont {D.}~\bibnamefont {Zhu}}, \bibinfo {author} {\bibfnamefont {L.}~\bibnamefont {Shao}}, \bibinfo {author} {\bibfnamefont {M.}~\bibnamefont {Yu}}, \bibinfo {author} {\bibfnamefont {R.}~\bibnamefont {Cheng}}, \bibinfo {author} {\bibfnamefont {B.}~\bibnamefont {Desiatov}}, \bibinfo {author} {\bibfnamefont {C.}~\bibnamefont {Xin}}, \bibinfo {author} {\bibfnamefont {Y.}~\bibnamefont {Hu}}, \bibinfo {author} {\bibfnamefont {J.}~\bibnamefont {Holzgrafe}}, \bibinfo {author} {\bibfnamefont {S.}~\bibnamefont {Ghosh}}, \bibinfo {author} {\bibfnamefont {A.}~\bibnamefont {Shams-Ansari}}, \emph {et~al.},\ }\href@noop {} {\bibfield  {journal} {\bibinfo  {journal} {Advances in Optics and Photonics}\ }\textbf {\bibinfo {volume} {13}},\ \bibinfo {pages} {242} (\bibinfo {year} {2021})}\BibitemShut {NoStop}%
\bibitem [{\citenamefont {Dorrer}\ and\ \citenamefont {Walmsley}(2002)}]{dorrer2002josab}%
  \BibitemOpen
  \bibfield  {author} {\bibinfo {author} {\bibfnamefont {C.}~\bibnamefont {Dorrer}}\ and\ \bibinfo {author} {\bibfnamefont {I.~A.}\ \bibnamefont {Walmsley}},\ }\href@noop {} {\bibfield  {journal} {\bibinfo  {journal} {JOSA B}\ }\textbf {\bibinfo {volume} {19}},\ \bibinfo {pages} {1019} (\bibinfo {year} {2002})}\BibitemShut {NoStop}%
\end{thebibliography}%
	%
	%
	%
\setcounter{equation}{0}
\setcounter{figure}{0}

\section*{Appendix}
\subsection{Quantum pulse gate (QPG) and its implementation for measuring the TPSC function}\label{qpg}
The proposed method in Sec.~\ref{TPSC function measurement} for measuring the TPSC function of pulses employs a quantum pulse gate (QPG). The implementation of this scheme is illustrated in Fig.~\ref{fig7}(a). In this scheme, we consider the pump profile is known prior to shaping. The QPG projects a single-photon-level input pulse, described by its complex spectral amplitude $f(\omega)$, onto any desired TF function decided by the pump TF mode $E_p(\omega_p)$, where $\omega_p$ is the pump frequency. The up-converted QPG output is centered at frequency $\omega_{out}$ and the energy conservation in this process fixes the relationship $\omega_{out}=\omega+\omega_p$. The total output counts $\eta$ at the output $\omega_{out}$ are given by following overlap integral 
\begin{equation}
	\eta \propto \left\lvert f(\omega) E^{*}_p(\omega_{out}-\omega)d\omega\right\rvert^2.
\end{equation}
This expression of $\eta$ is valid if both input and pump pulses are described by their complex spectral amplitudes. For partially coherent input pulses, $\eta$ in terms of the TPSC function $W(\omega_1,\omega_2)$ is given by
\begin{multline}
	\eta^{\theta}(\omega_{out}) \propto \iint W(\omega_1,\omega_2)E_p(\omega_{out}-\omega_1)\\
	\times E^*_p(\omega_{out}-\omega_2)d\omega_1d\omega_2.\label{I-qpg}
\end{multline}
\begin{figure*}[t!]
	\centering
	\includegraphics[scale=1]{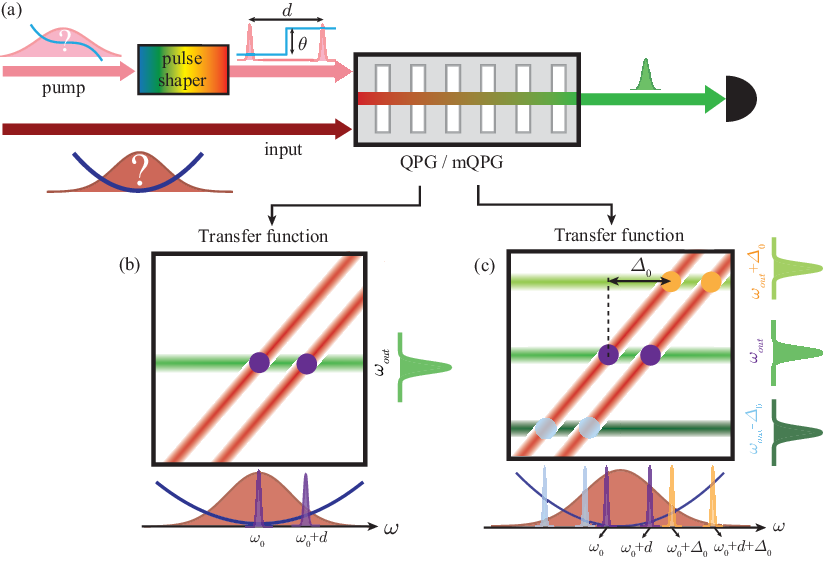}
	\caption{(a) Schematic of the proposed method for measuring TPDC function using a QPG and simultaneous characterization of input and pump pulses using mQPG. (b) Depicts the transfer function of QPG: the superposition of the frequency bin shape of pump up-converts input frequency bins at $\omega_0$ and $\omega_0+d$ into a single QPG output frequency $\omega_{out}$. The output counts provides resulting interference counts between the bins. (c) Depicts the transfer function: input frequency bins with separation $d$ for different shearing $\Delta$ up-converts into different mQPG output frequencies.}\label{fig7}
\end{figure*}

To measure the TPSC function $W(\omega_0,\omega_0+d)$, we need to interfere the complex spectral amplitudes $f(\omega_0)$ and $f(\omega_0+d)$ at frequencies $\omega_0$ and $\omega_0+d$ respectively, using a QPG. We start with a spectrally characterized pump pulse and shape it with a complex spectral amplitude $E_p(\omega_p)=\delta(\omega_p-\omega_p^{(0)})+e^{i\theta}\delta(\omega_p-\omega_p^{(0)}+d)$, where $\omega_p^{(0)}$ is the pump central frequency, $d$ is the bin separation and $\theta$ is the relative phase difference between the bins. As shown in Fig.~\ref{fig7}(b), the pump profile $E_p(\omega_p)$ up-converts a pair of spectral bins from the input to the same QPG output frequency $\omega_{out}$. As a result, the up-converted spectral bins interfere and the information is encoded in the total output counts of the QPG. 

The total count distribution $\eta^{\theta}(d)$ as a function of $d$ at the output $\omega_{out}$ can be written by substituting $E_p(\omega_p)$ into Eq.~(\ref{I-qpg}):
\begin{multline}
	\eta^{\theta}(d) = S(\omega_0)+S(\omega_0+d)+2\mathrm{Re}[W(\omega_0,\omega_0+d)] \\
	\times \cos\theta  +2\mathrm{Im}[W(\omega_0,\omega_0+d)]\sin\theta,\label{I-qpg_1_m}
\end{multline}
In the above expression, we replace $\omega_{p1,2}=\omega_{out}-\omega_{1,2}$ and define $\omega_0=\omega_{out}-\omega_p^{(0)}$. In order to extract $W(\omega_0,\omega_0+d)$, we measure $\eta^{\theta}(d;\omega_{out})$ at $\theta=0, \pi/2, \pi$, and $3\pi/2$ and obtain the real and imaginary parts of $W(\omega_0,\omega_0+d)$ 
\begin{equation}
	\mathrm{Re}[W(\omega_0,\omega_0+d)] \propto \eta^{\theta=0}(d) - \eta^{\theta=\pi}(d), 
\end{equation}
\begin{equation}
	 \mathrm{Im}[W(\omega_0,\omega_0+d)] \propto \eta^{\theta=\frac{\pi}{2}}(d) - \eta^{\theta=\frac{3\pi}{2}}(d).\label{I-qpg_2}
\end{equation}
By combining $\mathrm{Re}[W(\omega_0,\omega_0+d)]$ and $\mathrm{Im}[W(\omega_0,\omega_0+d)]$, we directly find $W(\omega_0,\omega_0+d)$. 
\begin{equation}
	W(\omega_0,\omega_0+d) = \mathrm{Re}[W(\omega_0,\omega_0+d)]+i \mathrm{Im}[W(\omega_0,\omega_0+d)].
\end{equation}

\subsection{Algorithm for simultaneously characterizing input and pump pulses}\label{mqpg algorithm_1}
In this case, we consider that input and pump pulses are described by their respective complex spectral amplitudes $f(\omega)$ and $g(\omega_p)$, where $\omega$ and $\omega_p$ are the input and pump frequencies. Our objective is to characterize the spectral phase profiles $\phi(\omega)$ and $\alpha(\omega_p)$ of input and pump respectively. For simplicity, we assume that the spectral amplitude of the pump pulse is uniform. 

As outlined in main manuscript Sec.~\ref{TPSC function measurement}, we extend the above TPSC measurement scheme to an mQPG, which introduces a spectral shear $\Delta$ between input and pump frequencies. Figure~\ref{fig7}(a) shows the illustration of this scheme with an mQPG with three output channels centered at $\omega_{out}+\Delta$, where $\Delta\in\{-\Delta_0,0,+\Delta_0\}$. Figure~\ref{fig7}(c) depicts that the output channel with frequency $\omega_{out}+\Delta$ enables the inteference of input frequencies $\omega_0+\Delta$ and $\omega_0+d+\Delta$ by still shaping the pump with the amplitude $E_p(\omega_p)=\delta(\omega_p-\omega_p^{(0)})+e^{i\theta}\delta(\omega_p-\omega_p^{(0)}+d)$. By measuring the count distribution $\eta(d;\Delta)$ for $\theta\in\{0,\pi/2,pi,3\pi/2\}$, we obtain the phase profile $\mu(d;\Delta)$ using
\begin{equation}
	\begin{aligned}
		\mu(d;\Delta) &= \mathrm{Arg}\left[\frac{\eta^{\theta=\frac{\pi}{2}}(d;\Delta)-\eta^{\theta=\frac{3\pi}{2}}(d;\Delta)}{\eta^{\theta=0}(d;\Delta)-\eta^{\theta=\pi}(d;\Delta)}\right]  \\
		&= \phi(\omega_0+d+\Delta)-\alpha(\omega^{(0)}_p-d)-\phi(\omega_0+\Delta).\label{mqpg1-ph}
	\end{aligned}
\end{equation}
This profile encodes both input and pump phase profiles. We retrieve the above phase profile for different $\Delta$ values. From these phase profiles, we first reconstruct the pump spectral phase profile by shifting the $\mu(d;\Delta)$ by $-\Delta$ and subtract it from $\mu(d;0)$

\begin{equation}
	{\Delta}\alpha(d)=\mu(d;0)-\mu(d-\Delta;\Delta) = \alpha(\omega^{(0)}_p-d+\Delta)-\alpha(\omega^{(0)}_p-d).\label{ph_diff_mqpg}
\end{equation}
The above profile is independent of the input spectral phase profile $\phi(\omega_0+d)$ and depends only on the pump phase difference profile $\alpha(\omega^{(0)}_p-d)-\alpha(\omega^{(0)}_p-d+\Delta_0)$. Notably, the SPIDER algorithm \cite{walmsley1996josab,iaconis1998optlett,dorrer2002josab} also provides the same phase difference profile while characterizing a single pulse. Thus, inspired by SPIDER and as a proof of principle, we impose the assumption that the pump phase profile $\alpha(\omega^{(0)}_p-d)$ is a polynomial function of the form $\alpha(\omega^{(0)}_p-d)=a\cdot(\omega^{(0)}_p-d)^2+ b\cdot(\omega^{(0)}_p-d)^3$, which is reasonable in pulse characterization experiments. In the experiment, we implement shearing values $\Delta\in\{-\Delta_0,+\Delta_0\}$, which yields two distinct ${\Delta}\alpha(d)$ profiles. A polynomial curve fitting, with a known $\Delta_0$, is then used on ${\Delta}\alpha(d)$ profiles to extract $\alpha(\omega^{(0)}_p-d)$. For small $\Delta_0$, as is the case for SPIDER, the phase difference profile in Eq.~(\ref{ph_diff_mqpg}) would directly map the derivative of $\alpha(\omega^{(0)}_p-d)$ \cite{dorrer2002josab}. Next, we obtain the input phase profile $\phi(\omega_0 + d)$ as
\begin{equation}
	\phi(\omega_0 + d) = \mu(d;0)-\alpha(\omega^{(0)}_p-d)+\phi(\omega_0). \label{mqpg3}
\end{equation}
Here, $\phi(\omega_0)$ introduces a constant offset to the reconstructed $\phi(\omega_0 + d)$ profile. Fig.~\ref{fig2} illustrates the steps of this phase retrieval scheme. Thus, by measuring the interference counts corresponding to different shearing values (at different mQPG output channels), we characterize the spectral phase profiles of both input and pump pulses. 

\subsection{Experimental details}\label{exp}
We first describe the setup for characterizing single-photon level, spectrally perfectly coherent pulses. A Ti:Sapphire pulsed laser of central wavelength 860 nm and repetition rate 80 MHz drives an OPO process to generate input pulses centered at 1545 nm. The residual 860 nm pulses serve as the pump and are directed to a in-house-built 4f-line pulse shaper to ``carve out" a superposition of frequency bins with a well-defined separation and relative phase. The bin width ranges from $30$ to $60$ GHz. A commercial wave shaper (Finisar 4000S) is used to   apply phase and amplitude profiles to produce custom-shaped coherent input pulses. After shaping, the input pulses are attenuated to mean photon of 0.1 per pulse using a neutral density filter. 

Both the input and pump pulses are sent to a periodically poled, titanium-indiffused LiNbO$_3$ QPG waveguide, 4 cm in length with a poling period of 4.32 $\upmu$m, operated at 433 K. The waveguide supports only the fundamental spatial mode for the input pulse, and appropriate care is taken to couple the pump pulse in the fundamental mode. The coupling efficiency for both input and pump pulses is approximately 70\% and 50\% respectively. The up-converted QPG output pulses, centered around 552 nm (543 THz), are separated from the residual input and pump pulses using a dichroic mirror. We detect the up-converted output counts using a time-of-flight (TOF) single-photon spectrograph with an effective resolution of 300 GHz. The spectrograph consist of a dispersive fiber, an avalanche photodiode (APD), and a time-to-digital converter (TDC). 
We keep the integration time per measurement point ranges between $3-7$ seconds . 
We note that due to the imperfections in QPG phase-matching, additional sidelobes of the phase-matching can reduce the QPG operation fidelity. To mitigate this, 
narrow band spectral filtering of the output photons is required prior to photon counting. Here, the TOF spectrograph implements both spectral filtering and photon counting over the filtered photons in a single device. Alternatively, one can use a narrow spectral filter alongside a single-photon detector to achieve the same goal and with a higher detection efficiency. 

For characterizing spectrally partially coherent quantum pulses, the combination of Ti:Sapphire pulsed laser and OPO is replaced by an integrated type-0 PDC source, realized in a 1-cm-long periodically pole titanium-indiffused LiNbO$_3$ waveguide operating at 443 K. The PDC process is pumped by a pulsed laser centered at 768 nm with a spectral bandwidth of 0.25 THz, generating partially coherent pulses centered at 1536 nm, which are sent to the QPG waveguide. 
A combination of narrow spectral filter and APD is used photon counting at the QPG output. The rest of the setup remains the same as the above. 

We now describe the setup for the simultaneous characterization of input and pump pulses. In this setup, we use an mQPG instead of a QPG. Using a Ti:Sapphire pulsed laser combined with an OPO, we generate spectrally coherent input pulses centered at 1545 nm. A in-house-built pulse shaper applies spectral phase profiles to the pump pulses, while a commercial waveshaper (Finisar 4000S) applies custom-shaped phase profiles to the input pulses. We attenuate the input pulses to a mean photon number of 1.0 per pulse.
The mQPG used in this experiment is a 4-cm-long titanium-indiffused LiNbO$_3$ waveguide operated at 433 K with a super-poling structure consisting of unpoled regions alternating to periodically poled regions with a poling period of 4.32 $\upmu$m. This poling structure generates three output frequency channels centered around 551.48 nm (543.62 THz), 552 nm (543 THz), and 552.62 nm (542.38 THz), corresponding to a shearing value of $\Delta_0/2\pi=$0.62 THz. The TOF spectrograph spectrally separates each channel and performs parallel measurements of total counts. This configuration can be replaced by combination of three spectral filters and single-photon detectors, which can boost the detection efficiency by up to two orders of magnitude.

\end{document}


\title{Supplementary Material: Frequency-bin interferometry for reconstructing electric fields with low intensity}

	\author{Abhinandan Bhattacharjee$^{\dagger}$, Laura Serino$^{\dagger}$, Patrick Folge$^{\dagger}$, Benjamin Brecht, and Christine Silberhorn}
	\affiliation{Integrated Quantum Optics Group, Institute for Photonic Quantum Systems (PhoQS), Paderborn University, Warburger Straße 100, 33098 Paderborn, Germany}
	\affiliation{$^{\dagger}$These authors contributed equally to this work.\\
		$^{*}$abhib@mail.uni-paderborn.de}
\maketitle
\section{Retrieval of spectral amplitude with an unknown pump pulse}
In this section, we present the steps for characterizing the spectral amplitude profile of the input pulse with an unknown pump pulse using a multi-output quantum pulse gate (mQPG).  
The input and pump pulses are described by their respective complex spectral amplitudes $f(\omega)$ and $g(\omega_p)$, where $\omega$ and $\omega_p$ are input and pump frequencies. 
In the proposed scheme, we introduce a spectral shear $\Delta$, to the input frequencies and interfere $\omega_0+\Delta$ and $\omega_0+d+\Delta$ by shaping the pump with a complex spectral amplitude amplitude $E_p(\omega_p)=\delta(\omega_p-\omega_p^{(0)})+e^{i\theta}\delta(\omega_p-\omega_p^{(0)}+d)$. This interference can be measured at an output frequency shifted by $\Delta$ relative to $\omega_{out}$ of a multi-output QPG. The resulting interference count distribution $\eta(d;\Delta)$ is given by: 
\begin{equation}
	\eta^{\theta}(d;\Delta) = \lvert{f(\omega_0+\Delta)g(\omega_p^{(0)})+f(\omega_0+d+\Delta)g(\omega_p^{(0)}-d)e^{i\theta}e^{i\alpha(\omega_p^{(0)}-d)}}\rvert^2.\label{I-qpg_4}
\end{equation}
We record $\eta^{\theta}(d;\Delta)$ at 0, $\pi/2$, $\pi$, and $3\pi/2$. These measurements yield the amplitude profile
\begin{equation}
	\begin{aligned}
		A(d;\Delta) &= |f(\omega_0)||g(\omega_p^{(0)})||f(\omega_0+d+\Delta)||g(\omega_p^{(0)}-d)|.\label{mqpg1-amp1}
	\end{aligned}
\end{equation}
The product $|f(\omega_0)||g(\omega_p^{(0)})|$ is a constant scaling factor, $K$. The above profile is simplified as the product of spectral amplitudes of the input and pump pulses. We note that the pump spectral amplitude $|g(\omega_0+d)|$ can be characterized by directly measuring the pump spectrum, $S(\omega_p)=|g(\omega_p)|^2$, using a spectrograph. From the measured spectrum, the input spectral amplitude can be retrieved as  
\begin{equation}
	|f(\omega_0+d+\Delta)| = \frac{1}{K} \frac{A(d;\Delta)}{\sqrt{S(\omega_p^{(0)}+d)}}.\label{mqpg1-amp2}
\end{equation}
%
%

Alternatively, the spectral amplitudes of both input and pump can be characterized simultaneously measuring $\eta^{\theta}(d;\Delta)$ at different values of $\Delta$  corresponding to each out channel of the mQPG. These measurements provide the amplitude profiles $A(d;-\Delta_0)$, $A(d;0)$, and $A(d;+\Delta_0)$ for $\Delta=-\Delta_0$, $0$, and $+\Delta_0$ respectively. The ratio 
\begin{equation}
	A_{\pm}(d) = \frac{A(d\pm\Delta_0;0)}{A(d;\pm\Delta_0)} = c \frac{g(\omega_p^{(0)}-d\mp\Delta_0)}{g(\omega_p^{(0)}-d)}.
\end{equation}
only depends on the pump amplitude ratios and independent of the input amplitude. 

For proof of principle, we assume that the pump has a Gaussian spectral amplitude of unknown bandwidth and center, which is a reasonable assumption in typical pulse characterization experiments. Using a curve-fitting on the $A_{\pm}(d)$ profiles for a known $\Delta_0$, we retrieve $g(\omega_p^{(0)}-d)$. The input amplitude profile $|f(\omega_0 + d)|$ then can be retrieved as
\begin{equation}
	|f(\omega_0 + d)| = K \frac{A(d;0)}{|g(\omega_p^{(0)}-d)|}, \label{mqpg3}
\end{equation}
where $K$ is a scaling constant. This retrieval approach is analogous to that of the simultaneous phase characterization of both input and pump pulses.

\section{Retrieval of spectral and phase profiles from the measured data}\label{pulsechar}
\begin{figure*}[t!]
	\centering
	\includegraphics[scale=1]{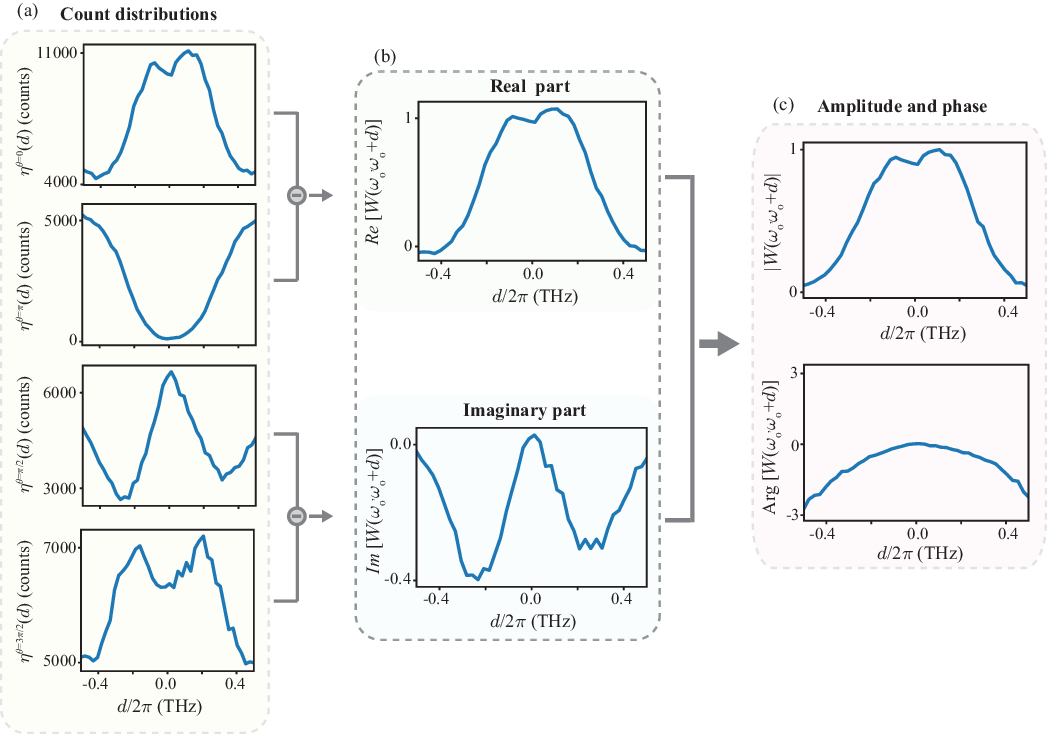}
	\caption{(a) Measured count distributions $\eta^{\theta}(d)$ at $\theta=0,\pi,\pi/2$, and $3\pi/2$. (b) Real and imaginary profiles of $W(\omega_0,\omega_0+d)$. (c) Amplitude $|W(\omega_0-d,\omega_0+d)|$ and Arg$[W(\omega_0-d,\omega_0+d)]$ profiles of $W(\omega_0-d,\omega_0+d)$.
	}\label{fig_1}
\end{figure*} 

In this section, we outline the procedure for retrieving the spectral amplitude and phase profiles from the measured data. The analysis focuses on a spectrally perfectly coherent input pulse, as depicted in the first column of Fig. 4(a) in the main manuscript. Figure~\ref{fig_1} shows the measured count distributions $\eta^{\theta}(d)$ at $\theta=0,\pi/2,\pi, 3\pi/2$. The difference in count distributions yield the real and imaginary parts of the two-point spectral correlation (TPSC) function $W(\omega_0,\omega_0+d)$ as shown in Fig.~\ref{fig_1}(b). By combining them, we retrieve the spectral amplitude $|W(\omega_0,\omega_0+d)|$and phase Arg$[W(\omega_0,\omega_0+d)]$ profiles as shown in Fig.~\ref{fig_1}(c).  

\section{Characterizing partially coherent pulses with tunable time-frequency coherence}
 
In this section, we focus on characterizing the spectral coherence profile of different spectrally partially coherent single-photon-level pulses generated by incoherently mixing perfectly coherent pulses from an OPO in post-processing. The incoming pulses are shaped into different Hermite-Gaussian (HG) modes $\rm{HG}_n(\omega)$ and the corresponding TPSC functions, $W_n(\omega_0-d,\omega_0+d)$, are measured. To realize a partially coherent pulse in the post-processing, we incoherently add the measured $W_n(\omega_0-d,\omega_0+d)$ data with weights $\lambda_n$, forming $W(\omega_0-d,\omega_0+d)=\sum_n \lambda_n W_n(\omega_0-d,\omega_0+d)$ corresponding to a partially coherent pulse. In experiment, we keep the input mean photon number at 0.1 photons per pulse. The total data accumulation time for each partially coherent pulse takes around 120 minutes.

\begin{figure*}[t!]
	\centering
	\includegraphics[scale=1]{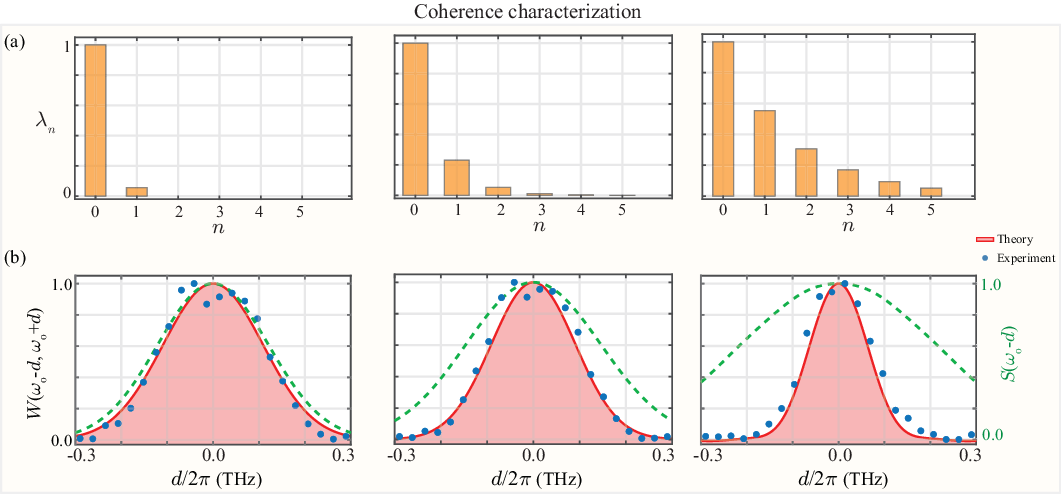}
	\caption{(a) Mode distribution $\lambda_n$ for different incoherent mixtures. (b) $W(\omega_0-d,\omega_0+d)$ profiles for different incoherent mixtures alongside theoretical predictions. The green dashed curves represent the expected spectrum $S(\omega_0-d)$ for each incoherent mixture.
	}\label{fig_W}
\end{figure*}

Figure~\ref{fig_W}(a) shows the mode distribution $\lambda_n$ as a function of the mode index $n$ for different incoherent mixtures. Here, we are interested in the relative proportion of individual HG pulses; an appropriate normalization of $\lambda_n$ is not necessary. Figure~\ref{fig_W}(b) show the corresponding measured $W(\omega_0-d, \omega_0+d)$ along with the theoretical predictions. The dashed green curves in Fig.~\ref{fig_W}(b) represent the expected spectrum $S(\omega_0-d)$ for corresponding to each partially coherent pulses. Furthermore, we see that $W(\omega_0-d, \omega_0+d)$ becomes narrower with the increases in the number of HG modes in the incoherent mixture, indicating a decrease in the spectral coherence. A quantitative comparison between theoretical and experimental results yields similarity $S \approx$ 99\% in both cases, demonstrating the accurate coherence characterization capability of this scheme.   

\section{Time-frequency characteristics of type-0 parametric down-conversion (PDC) process}
In this section, we describe the time-frequency (TF) properties of the type-0 PDC process and explain how it generates spectrally partially coherent pulses. A Gaussian pump pulse, centered at frequency $\omega^{(0)}_{pump}$ with a spectral bandwidth $\sigma_p$, drives the type-0 PDC process, producing signal and idler photon pairs centered at frequencies $\omega^{(0)}_{s}$ and $\omega^{(0)}_{i}$, respectively. The energy conservation condition of this process imposes the relationship $\omega_{pump}=\omega_{s}+\omega_{i}$, where, $\omega_{pump}$, $\omega_s$, and $\omega_i$ represent the frequencies of pump, signal, and idler photons, respectively,

The joint spectral amplitude (JSA), which fully describes the complete TF structure of signal-idler pair, is expressed as   
\begin{figure*}[b!]
	\centering
	\includegraphics[scale=0.95]{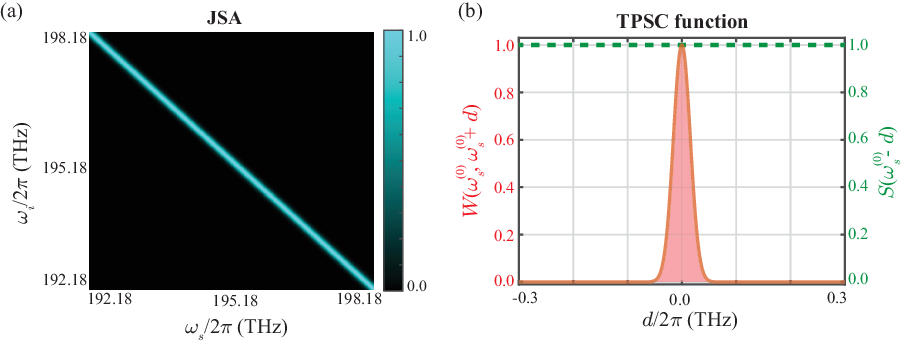}
	\caption{(a) Joint spectral amplitude (JSA) of our type-0 PDC source. (b) TPSC function  $W(\omega^{(0)}_s,\omega^{(0)}_s+d)$ of signal evaluated from the above JSA. The corresponding spectrum (green dashed line) $S(\omega_0-d)$ is also plotted for comparison.
	}\label{fig_pdc}
\end{figure*} 
\begin{figure*}[t!]
	\centering
	\includegraphics[scale=1.0]{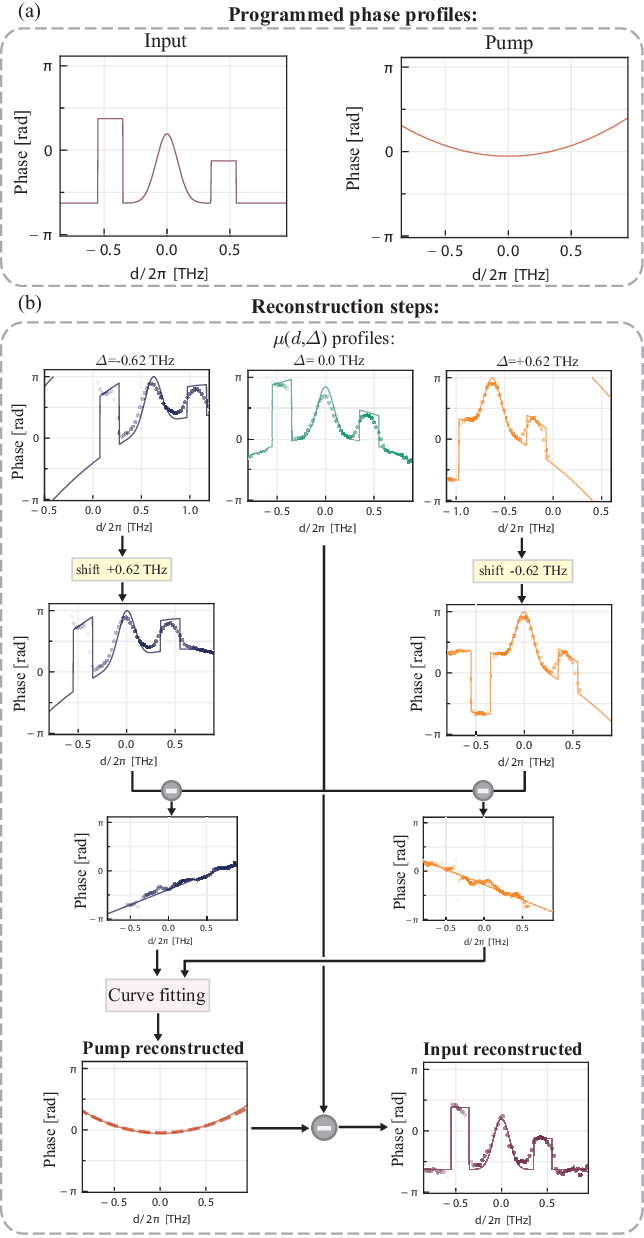}
	\caption{(a) Spectral phase profiles programmed by input and pump pulse shapers. (b) Steps for reconstructing the spectral phase profiles of input and pump pulses from the measured data. The solid curves represent theoretical predictions and dots represent the measured and reconstructed profiles.
	}\label{fig_mqpg}
\end{figure*} 
\begin{equation}
	\psi(\omega_s,\omega_i) = N \exp\left[-\frac{\left(\omega_s+\omega_i-\omega^{(0)}_s-\omega^{(0)}_i\right)^2}{4\sigma_p^2}\right] \Phi(\omega_s,\omega_i), \label{type-0_PDC}
\end{equation}
where, $\Phi(\omega_s,\omega_i)$ represents the phase-matching function determined by the material properties of the LiNbO$_3$ waveguide used in our experiment to realize the type-0 PDC process. Figure~\ref{fig_pdc}(a) shows the JSA $\psi(\omega_s,\omega_i)$ corresponding to our type-0 PDC setup, showcasing the strong spectral correlation between signal and idler photons. Here, we see that signal and idler photons are spectrally identical. 

The JSA $\psi(\omega_s,\omega_i)$ can be decomposed into Schmidt modes $\{f_n(\omega)\}$ as    
\begin{equation}
	\psi(\omega_s,\omega_i) = \sum_n \lambda_n f_n(\omega_s)f_n(\omega_i), \label{type-0_PDC-schimdt}
\end{equation}
where, $\lambda_n$ is the Schmidt coefficient correspond to each Schmidt mode $f_n(\omega)$. To determine the TF structure of the signal photons alone, we trace out the idler photons from the JSA. This is described by the TPSC function $W(\omega_s,\omega'_s)$ 
\begin{equation}
	W(\omega_s,\omega'_s) = \int \psi(\omega_s,\omega_i)\psi^*(\omega'_s,\omega_i)d\omega_i=\sum_n |\lambda_n|^2 f_n(\omega_s)f^*_n(\omega'_s), \label{type-0_PDC-schimdt}
\end{equation}
%
This TPSC function represents an incoherent mixture of Schmidt modes $\{f_n(\omega_s)\}$, weighted by $\{\lambda_n\}$. This incoherent mixture leads to the generation of spectrally partially coherent signal pulses. Figure~\ref{fig_pdc}(b) compares the one-dimensional TPSC function $W(\omega_s^{(0)}\omega_s^{(0)}-d)$ with the spectrum $S(\omega_s^{(0))}-d)$. The significantly narrower width of the TPSC function, relative to the spectrum, highlights the low spectral coherence of the signal pulses.

\section{Reconstruction of spectral phase profiles of input and pump pulses}
Figure~\ref{fig_mqpg}(a) shows the spectral phase profiles of input and pump pulses programmed through respective pulse shapers that we want to characterize. Figure~\ref{fig_mqpg}(b) shows the reconstruction steps. First, following the methodology described in Sec.~\ref{pulsechar}, we reconstruct the phase profiles $\mu(d;\Delta)$ as shown in Fig.~\ref{fig_mqpg}(b) (top row) by measuring count distributions at different mQPG output channels characterized by $\Delta=-0.62, 0,$ and $+0.62$ THz alongside the corresponding theoretical predictions.

To retrieve the input and pump phase profiles, we use the algorithm illustrated in Fig.2 of the main text. Specifically, we introduce spectral shifts of $+0.62$ THz and $-0.62$ THz to $\mu(d;\Delta = -0.62)$ and $\mu(d;\Delta = +0.62)$, respectively. By subtracting these shifted profiles from $\mu(d;0)$ and then employing a curve-fitting algorithm, we retrieve the pump phase profile, $\alpha(\omega^{(0)}_p-d)$, which is shown in Fig.~\ref{fig_mqpg}(b) (last row left figure). Next, we subtract the obtained pump profile $\alpha(\omega^{(0)}_p-d)$ from $\mu(d;0)$ to retrieve the input phase profile, see last row right plot in Fig.~\ref{fig_mqpg}(b). This data set is presented in the main text in Fig.6(a) and (b). 

\bibliographystyle{apsrev4-2}
\bibliography{ref}